\documentclass[twocolumn,epjc3]{svjour3}
\RequirePackage{amsmath}
\RequirePackage{amssymb}
\RequirePackage{flushend}
\RequirePackage{graphicx}
\RequirePackage[numbers,sort&compress]{natbib}

\journalname{Eur. Phys. J. C}
\newcommand{\be}{\begin{equation}}
\newcommand{\ee}{\end{equation}}

\newcommand{\no}{\nonumber\\}
\newcommand{\ba}{\begin{eqnarray}}
\newcommand{\ea}{\end{eqnarray}}
\newcommand{\bg}{\begin{multline}}
\newcommand{\eg}{\end{multline}}

\newcommand{\eint}{\int_{\sigma_1}^{\infty} dE\, }
\def\gl#1{(\ref{#1})}
\def\tr#1{\mbox{tr}\left\{#1\right\}}

\newcommand{\la}[1]{\label{#1}}

\begin{document}

\title
{Spontaneous parity violation in extreme conditions:
an effective lagrangian analysis}
\author{Alexander A. Andrianov\thanksref{e1,addr1,addr2}
    \and
    Vladimir A. Andrianov\thanksref{e2,addr1}
    \and
    Domenec Espriu\thanksref{e3,addr2}
}

\thankstext{e1}{e-mail: andrianov@icc.ub.edu}
\thankstext{e2}{e-mail: v.andriano@rambler.ru}
\thankstext{e3}{e-mail: espriu@icc.ub.edu}

\institute{V.A. Fock Department of Theoretical Physics, Saint-Petersburg State University, ul. Ulianovskaya,1, 198504 St. Petersburg, Russia\label{addr1}
\and
Departament d’Estructura i Constituents de la Mat\'eria and Institut de Ci\'encies del Cosmos (ICCUB) Universitat de Barcelona, Mart\'i Franqu\`es, 1, E08028
Barcelona, Spain\label{addr2}
}

\date{Received: date / Accepted: date}

\maketitle

\begin{abstract}
We investigate how  large baryon densities (and possibly high temperatures) may induce
spontaneous parity violation in the composite meson sector of vector-like gauge theory (presumably QCD or techni-QCD) . The analysis at intermediate energy scales is done by using an extended $\sigma$-model
lagrangian that includes two scalar and two pseudoscalar multiplets and fulfills low-energy
constraints for vector-like gauge theories. We elaborate on a novel mechanism of parity breaking
based on the interplay between lightest and heavier meson condensates, which therefore
cannot be realized in the simplest $\sigma$ model.  The results
are relevant for an idealized homogeneous and infinite nuclear (quark or techniquark) matter where the
influence of density can be examined with the help of a constant chemical potential.
The model is able to describe satisfactorily the first-order phase transition
to stable nuclear matter, and predicts a second-order phase transition to a state where parity
is spontaneously broken. We argue that the parity breaking phenomenon is quite generic when a large enough
chemical potential is present. Current quark masses are explicitly taken into account in this work and
shown not to change the general conclusions.
\end{abstract}


\section{\label{sec:intro}Introduction}
Emergent parity violation for sufficiently large values of the baryon chemical potential
(and/or temperature) has been attracting much interest during several decades (see reviews \cite{picon}). Yet the reliable prediction of  parity violation effects has not been done from the first principles. In our work we investigate how  large baryon densities (and possibly high temperatures) may induce spontaneous parity violation in the composite meson sector of vector-like gauge theory (presumably QCD or techni-QCD). The analysis is performed by using an extended $\sigma$-model lagrangian that includes two scalar and two pseudoscalar multiplets and fulfills low-energy constraints. This is a model inspired by but not exactly equivalent to QCD as its coupling constants are taken as empirical parameters to be measured in meson physics.

At finite baryon density pion condensation is conjectured
in nuclear physics long ago in \cite{migdal} and it seems to be a plausible possibility which however cannot be proved in simple models describing pion-nucleon
interactions.  In this paper this long-standing idea will in fact be vindicated.

In this paper we shall attempt to explore the interesting issue of parity
breaking employing effective lagrangian techniques, useful to explore the range of nuclear
densities where the hadron phase still persists and quark percolation does not occur yet.
Our effective lagrangian is a realization of the generalized
linear $\sigma$ model, but including the two lowest lying resonances in each
channel, those that are expected to play a role in this issue. This seems to be the
minimal model where the interesting possibility of parity breaking can be realized. Namely,
condensation of one of the pseudoscalar fields can arise on the background of two-component scalar
condensate so that the chiral constant background cannot be rotated away by transformation of
two complex scalar multiplets preserving space parity. The use of
effective lagrangians is also crucial to understand how would
parity breaking originating from a finite baryon density eventually reflect in hadronic physics.

A pre-QCD attempt to describe
two multiplets of scalar and pseudoscalar mesons was done in \cite{bars} with
a reduced set of operators and a chiral symmetry breaking (CSB) pattern not quite compatible with QCD.
We have been basically inspired by our previous works on extended quark models
\cite{edt,ava,aet} where two different schemes with linear and non-linear
realization of chiral symmetry were adopted to incorporate heavy pions
and scalar mesons within an effective quark model with quark self-interactions.
 A certain resemblance can be also found with the model \cite{schech} where two $SU(3)_F$
multiplets have been associated with two-quark and four-quark meson states
although we do not share the assumption in \cite{schech} concerning the dominance of four-quark component in
radially excited mesons. The model in \cite{giacosa} is also of relevance in studying of vacuum for
extended $\sigma$ models.

The present work is an extension of the preliminary results concisely reported in  \cite{anes,aaes} which includes corrections beyond the chiral limit,
linear in the current quark masses.
We will give also the qualitative explanation of a possible origin of the model, provide the detailed proofs of several statements that were only enunciated
in  \cite{anes,aaes} and derive a number of new thermodynamic relations for finite temperatures and chemical potentials.

The paper is organized as follows. In section 2 the bosonization of QCD quark currents
in the color-singlet sector is discussed and the ingredients of
the generalized $\sigma$ model are indicated. In Section 3 we introduce the $\sigma$ model
with two multiplets of isosinglet scalar and isotriplet pseudoscalar fields.
The effective potential for two multiplets of scalar
and pseudoscalar mesons is obtained and the mass-gap equations and second variations at the minima are derived.
In Section 4  the existence of a region in the coupling constant space is proven where there are four minima of the effective potential playing the crucial role in realization of stable baryon matter via a first order phase transition.
In Section 5 we shall introduce the
finite chemical potential and temperature and see how they modify the effective theory and
the vacuum state. Temperature and baryon chemical potential appear through the one-quark
loop free energy.
In sections 6 we investigate the emergence of spontaneous parity breaking (SPB) phase. The mass gap
equations and critical lines for the parity breaking phase transition are derived and corrections
beyond the chiral limit are taken into account to the leading order in quark masses. In Section 7 it is established
that the transition to the SPB phase is of second order.  In Section 8 the kinetic
terms are considered in order to determine the physical masses of scalar and pseudoscalar mesons and extract some physical consequences. In particular, it is proven that for massive quarks only three massless states characterize the SPB phase transition. In SPB phase the masses of four light pseudoscalar  states  are obtained.
We notice however that in the SPB phase strictly speaking there are no genuine scalar or pseudoscalar states
as each of massive states can equally well decay into two and three (pseudo)scalars.
Section 9 is devoted to a description of nuclear matter and the approach to the condensation point of stable baryon matter. In order to describe adequately the saturation point transition to stable baryon matter
we supplement the effective lagrangian with an $\omega$ meson coupling to the isosinglet quark current
which influences the repulsive part
of nuclear forces\cite{walec} and thereby supports the formation of stable nuclear matter.
We obtain a first-order phase transition at the saturation point.  In Section 10 we attempt to confront the on-set of empirical constants of two-multiplet model with meson and nuclear matter phenomenology.
We summarize our findings  in the conclusions Section. In Appendix A we prove that the
chiral collapse affecting the simplest $\sigma$ models and/or the one-multiplet Nambu-Jona-Lasinio (NJL)
models \cite{buballa1} does not occur in our two-multiplet model. In Appendix B the proposal for description of (in)compressibilities in the mean-field approach is formulated with the help of matching quark and nuclear matters.

The range of intermediate nuclear densities where our effective lagrangian could be used
is of high interest as they may be reached in both compact
stars \cite{grei} and heavy-ion collisions \cite{heavyion}. Its relevance can be qualitatively
motivated by the fact that at substantially larger densities typical distances between
baryons are shrinking considerably
and meson excitations with Compton wave lengths much shorter than the pion wave length start playing an
important role. Can the spontaneous parity breaking be realized in heavy ion collisions or in neutron stars?
In order to answer this question we might appeal
to lattice QCD for help and in fact this possibility has been studied intensively for quite some time
\cite{philip,lombardo,steph}. However the lattice results for sufficiently large values of
the baryon chemical are not known quantitatively and rigorously yet.

It is worth to mention some previous studies dealing with the problem of strong interactions
at zero temperature and finite chemical potential: depending on a value of nuclear density, a
variety of methods  are involved from using meson-nucleon
\cite{picon,bary} or quark-meson\cite{grei,sigqu}
lagrangians  for low-dense nuclear matter  to models of the NJL type\cite{NJL}
for high-dense quark matter\cite{scon}.
Although the issue of SPB in hadronic
phase has been touched upon in the pion-nucleon
theory\cite{picon,bary,takah} and in NJL models \cite{ebert} the reliability  of the models
used is not quite clear for intermediate nuclear densities. The reason is discussed in the
next Section: they are not rich enough to explore the subtle phenomenology involved.

More recently the phenomenon of parity breaking was assumed to be present in meta-stable nuclear bubbles
with non-zero axial charge generated by nontrivial topological charge in hot nuclear matter \cite{kharzeev} and/or in the presence of a strong background magnetic fields \cite{kharmacl,polikarpov}. It was also shown \cite{aaep} that the associated axial chemical potential causes a distortion of the energy spectrum of photons and vector particles
($\rho$ and $\omega$ mesons) due to a Chern-Simons term that is generated. In addition scalar and
pseudoscalar mesons get a momentum dependent effective mass \cite{aepl}. However this phenomenon
is theoretically somewhat different in its origin to the previous one and it will not be discussed in the present paper.
\section{\label{qcd-bosonization} Bosonization of vector-like gauge theories in the color-singlet sector}
In order to elaborate an effective lagrangian for composite meson states starting from QCD or QCD-like theories
we revisit the properties of color-singlet (quasi)local quark currents in the
vacuum with spontaneously broken chiral symmetry. This phenomenon  emerges due to a non-zero value
of quark condensate $\langle \bar q q \rangle
$ and can be associated to the CSB scale $
\Lambda$ (in QCD it is presumably $ \sim 1 GeV$). This CSB due to quark condensation
makes the quark bilinears to be interpolating operators for meson fields (in the limit of large number of
colors). In particular, the scalar and pseudoscalar quark densities effectively describe the
creation or annihilation of scalar and pseudoscalar mesons
\begin{eqnarray}
\bar q q(x) &\simeq& \Lambda^2 \sum\limits_{l=1}^\infty Z^{(1)}_{l}\sigma_l(x);\no
\bar q\gamma_5 \tau^a q(x) &\simeq& \Lambda^2
\sum\limits_{l=1}^\infty Z^{(1)}_{l}\pi^a_l(x), \label{interp}
\end{eqnarray}
where the normalized meson fields $\sigma_l,\pi^a_l $ describe the families of
resonances with the same quantum numbers but increasing masses
(radial Regge trajectories)\cite{afo} and the set of normalization constants $Z^{(1)}_{l}$ is introduced.
The constituent quark fields are denoted as $\bar q, q$ and $\tau^a, a=1,2,3$ stand for Pauli
matrices. At this stage we consider the chiral limit of zero current quark masses. Accordingly the
global chiral covariance of quark operators \gl{interp} is transmitted to the set of boson operators
leading to an equal normalization of scalar and pseudoscalar fields. This is a basic framework
of linear sigma models \cite{gellmann} (see next Section). On the other hand the CSB phenomenon must be transmitted to condensation of scalar fields so that the quark condensate is interpolated by v.e.v.'s $\langle\sigma_l\rangle$ of scalar fields,
\begin{equation}
\langle\bar q q\rangle \simeq \Lambda^2 \sum\limits_{l=1}^\infty Z^{(1)}_{l} \langle\sigma_l\rangle ,
\end{equation}
which represents the condition on the choice of potential in a QCD motivated sigma model.
 In this paper we restrict ourselves
with consideration of two light flavors related to $u,d$ quarks and therefore the
approximate chiral symmetry of the quark sector is \mbox{$SU(2)_L \times SU(2)_R$}.

Keeping in mind confinement we have retained  in \gl{interp} only the
one-resonance states as leading ones while being aware of that the total saturation of quark
currents includes, of course, also multi-resonance states. Thus we use the
large-$N_c$ approach where resonances behave like true elementary particles
with zero widths and multi-resonant states can be neglected.

Let us comment a bit more on the previous relation. On the left-hand side one sees an operator of
canonical dimension 3 whereas on the right-hand side one finds field operators of canonical dimension 1.
This drastic change in dimensions is a consequence of CSB and it modifies the
dimensional analysis of what must be included into an effective
lagrangian. More exactly, in order to replace the non-perturbative regime of QCD at low and
intermediate energies by a hadron effective lagrangian one has to apply this
dimensional counting in the CSB phase \cite{ava} to all possible combinations of color-singlet
operators arising in the chiral
expansion in inverse powers of the CSB scale $\Lambda$.

To be specific the chiral invariant local operators playing the leading role in the low-energy
effective lagrangian for meson dynamics are
\begin{eqnarray}
&&\frac{1}{\Lambda^2}\left[(\bar q q)^2  -
\bar q\gamma_5 \tau_a q \bar q\gamma_5 \tau^a q\right]\no&&\simeq \Lambda^2
\sum\limits_{l,m=1}^\infty Z^{(2)}_{lm}[\sigma_l \sigma_m + \pi^a_l\pi_{a,m}]
;\nonumber\\
&&\frac{1}{\Lambda^8}[(\bar q q)^2  -
\bar q \gamma_5 \tau_a q\bar q\gamma_5 \tau^a q]^2 \no&&\simeq
\sum\limits_{l,m,n,r=1}^\infty Z^{(4)}_{lmnr}[\sigma_l \sigma_m +
\pi^a_l\pi_{a,m}][(\sigma_n \sigma_r + \pi^a_n \pi_{a,r}]
;\nonumber\\
&&\frac{1}{\Lambda^4}[\partial_\mu(\bar q q)\partial^\mu(\bar q q)  -
\partial_\mu (\bar q\gamma_5 \tau_a q) \partial^\mu (\bar q\gamma_5 \tau^a
q)] \nonumber\\
&&\simeq  \sum\limits_{l,m=1}^\infty \tilde Z^{(2)}_{lm}[\partial_\mu \sigma_l
\partial^\mu\sigma_m + \partial_\mu\pi^a_l\partial^\mu\pi_{a,m}]
,
\label{boso}
\end{eqnarray}
where the matrices $Z^{(2)}_{lm}, Z^{(4)}_{lmnr},\tilde Z^{(2)}_{lm} $ must be
symmetric under transposition of indices in order to provide global chiral invariance. The superscript numbers indicate the powers of interpolating meson fields. The terms quadratic in
scalar fields must trigger an instability in the potential that leads to CSB in
the effective meson theory due to condensation of scalar fields, $\langle\sigma_l\rangle \not= 0$ for some $l$ (see. e.g. \cite{ava,schech,buballa1}).

The above set of operators is not complete and can be extended with the help of form factors
that are polynomials in derivatives \cite{ava}.
For example, using the same CSB scale $\Lambda$ one can add into the effective quark lagrangian the
vertices built of the elements
\begin{eqnarray}
\bar q\frac{\overleftrightarrow\partial^{2k}}{\Lambda^{2k}} q(x) &\simeq& \Lambda^2
\sum\limits_{l=1}^\infty Z^{(1),k}_{l}\sigma_l(x);\nonumber\\
\bar q\gamma_5 \tau^a \frac{\overleftrightarrow\partial^{2k}}{\Lambda^{2k}} q(x) &\simeq& \Lambda^2
\sum\limits_{l=1}^\infty Z^{(1),k}_{l}\pi^a_l(x),\no
\overleftrightarrow\partial^{2} &\equiv& \frac14 (\overrightarrow\partial_\mu-\overleftarrow\partial_\mu) (\overrightarrow\partial^\mu-\overleftarrow\partial^\mu)
, \la{quasi1}
\end{eqnarray}
which may give  numerically comparable contributions for several $k$ \cite{ava}.

\section{\label{sigma-model} Generalized sigma-model}
\subsection{Effective potential for two multiplets of scalar and pseudoscalar fields}
The simplest hadronic effective theory is the linear sigma-model of Gell-Mann and Levy \cite{gellmann},
which contains a multiplet of the lightest scalar $\sigma$ and pseudoscalar $\pi^a$
fields. Spontaneous CSB  emerges due to a non-zero value
for $\langle \sigma\rangle  \sim \langle \bar q q \rangle /\Lambda^2,\ \Lambda \sim 4\pi F_\pi$ with $F_\pi$ being a weak pion decay coupling constant. Current algebra techniques
indicate that in order to relate this model to QCD one has to choose a real condensate for the scalar density,
with its sign opposite to current quark masses, and avoid any parity breaking due to a v.e.v. of the
pseudoscalar density. The introduction of a chemical potential does not change the phase of the
condensate and therefore does not generate any parity breaking. This is just fine because in normal
conditions parity breaking does not take place in QCD.
However, if two different scalar fields condense with a relative phase between
the two v.e.v.'s the opportunity of spontaneous parity breaking may arise.

Let us consider a model with two multiplets of scalar ( $\tilde{\sigma}_j$) and
pseudoscalar ( $\tilde{\pi}^a_j$ ) fields
\begin{equation}
H_j = \tilde{\sigma}_j {\bf I}
+ i \hat\pi_j, \quad j = 1,2;\quad H_j H_j^\dagger = (\tilde{\sigma}^2_j +
(\tilde{\pi}^a_j)^2 )
{\bf I} , \label{linpar}
\end{equation}
where ${\bf I}$ is an identity $2\times 2$ matrix and $\hat\pi_j \equiv \tilde{\pi}^a_j \tau^a$ with $\tau^a$ being a set of Pauli matrices.
We shall deal with a scalar system globally symmetric respect to $SU(2)_L \times SU(2)_R$
rotations in the exact chiral limit and next consider the soft breaking of chiral
symmetry by current quark masses. We should think of
these two chiral multiplets as representing the two lowest-lying radial states
for a given $J^{PC}$. Of course one could add more multiplets, representing higher radial and spin excitations,
to obtain a better description of QCD, but the present model, without being completely realistic,
already possesses all the necessary ingredients to study SPB.  Inclusion of
higher-mass states would be required at substantially larger baryon densities when typical
distances between baryons are shrinking considerably and meson excitations
with Compton wave lengths much shorter than the  pion wave length start playing an important role.

Let us define the effective potential of this generalized sigma-model. First
we write the most general Hermitian potential at zero $\mu$,
\begin{eqnarray}
 V_{\mbox{\rm eff  }}&=& \frac12 \tr{- \sum_{j,k=1}^2
H^\dagger_j \Delta_{jk} H_k + \lambda_1 (H^\dagger_1 H_1)^2 \right.\nonumber\\ && \left.+ \lambda_2
(H^\dagger_2 H_2)^2+
\lambda_3 H^\dagger_1 H_1 H^\dagger_2 H_2 \right.\nonumber\\ && \left.+ \frac12
\lambda_4 (H^\dagger_1 H_2
H^\dagger_1 H_2 + H^\dagger_2 H_1 H^\dagger_2 H_1)\right.\nonumber\\ && \left. + \frac12 \lambda_5
(H^\dagger_1 H_2 + H^\dagger_2
H_1) H^\dagger_1 H_1  \right.\nonumber\\ &&\left.+ \frac12 \lambda_6 (H^\dagger_1 H_2 +
H^\dagger_2 H_1)
H^\dagger_2 H_2  } + {\cal O}(\frac{|H|^6}{\Lambda^2}), \label{effpot1}
\end{eqnarray}
which contains 9 real constants $\Delta_{jk}, \lambda_A; \ A= 1,\ldots,6.$ However this
set of constants can be reduced (see sect.\ref{reduc}). QCD bosonization rules in the large $N_c$
limit prescribe $\Delta_{jk}\sim \lambda_A \sim N_c$. The neglected terms will be suppressed
by inverse power of the CSB scale $\Lambda \sim 1$ GeV. If we assume the v.e.v. of $H_j$ to be of the order of
the constituent mass $0.2 \div 0.3$ GeV,  it is reasonable to neglect these terms.

One could add five more  terms (breaking parity manifestly): an imaginary part
of $\Delta_{12}$ with an operator
\begin{equation}
i \tr{(H^\dagger_1 H_2 - H^\dagger_2 H_1)},\label{option1}
\end{equation}
and four more operators
\begin{eqnarray}  &&i \tr{H^\dagger_1 H_2 H^\dagger_1 H_2 - H^\dagger_2 H_1 H^\dagger_2
H_1}, \nonumber\\ &&i \tr{(H^\dagger_1
H_2 - H^\dagger_2 H_1) H^\dagger_1 H_1 },\nonumber\\ && i \tr{(H^\dagger_1 H_2 -
H^\dagger_2 H_1) H^\dagger_2
H_2};\nonumber\\ && i\left[\Big( \tr{H^\dagger_1 H_2 }\Big)^2 -\Big( \tr{H_1 H_2^\dagger }\Big)^2\right].\label{morevert}\end{eqnarray}
But for the scalar multiplets \gl{linpar} in $SU(2)_L \times SU(2)_R$ representation these operators identically vanish (see below).

There are also two operators with two disconnected traces which seem to complete the full set of operators
\begin{eqnarray}
&&\Big( \tr{H^\dagger_1 H_2 }\Big)^2 +\Big( \tr{H_1 H_2^\dagger }\Big)^2;\quad\nonumber\\ && \tr{H^\dagger_1 H_2 }\tr{H_1 H_2^\dagger } . \label{option2}
\end{eqnarray}
However for the scalar multiplets \eqref{linpar} they are not independent and can be expressed as the linear combination of operators with constants $\lambda_3,\lambda_4$ in \gl{effpot1}. The proofs of above statements can be easily done with the help of so called chiral parameterization. Namely, one can use the global invariance of the model to factor
out the Goldstone boson fields with the help of the chiral parameterization
\begin{eqnarray}
&&H_1 (x) = \sigma_1 (x) \xi^2(x) ;\nonumber\\ && H_2 (x) = \xi
(x)\Big(\sigma_2 (x) +i
\hat\pi_2 (x)\Big)\xi (x) ;\no &&\xi \equiv \exp\left(i\frac{\pi_1^a
\tau_a}{2F_0}\right)\nonumber\\ && = \cos\left(\frac{\sqrt{(\pi_1^a)^2}}{2F_0} \right)+i \frac{\pi_1^a\tau_a}{\sqrt{(\pi_1^a)^2}}\sin\left(\frac{\sqrt{(\pi_1^a)^2}}{2F_0}\right),
 \label{chipar}
\end{eqnarray}
which differs from eq. \gl{linpar} in notation. The constant $F_0$ is related to the bare pion decay constant and will be defined later when the kinetic terms are normalized. This kind of parameterization
preserves the parities of $\sigma_2 (x)$ and $\hat\pi_2$ to be even and odd respectively
in the absence of SPB. Then the contribution of the four additional operators \gl{option1},\gl{morevert}
vanishes identically, whereas the operator \gl{option2} turns out to be a combination of
operators with constants $\lambda_{3,4}$. Finally the potential \gl{effpot1} is further simplified to
\begin{eqnarray}  &&\!\!\!\!V_{\mbox{\rm eff  }}= - \sum_{j,k=1}^2
\sigma_j \Delta_{jk}
\sigma_k - \Delta_{22} (\pi_2^a)^2 \\ &&\!\!\!\!+  \lambda_2  \Big((\pi_2^a)^2\Big)^2   +
\Big((\lambda_3 - \lambda_4) \sigma_1^2 + \lambda_6 \sigma_1 \sigma_2 +
2\lambda_2 \sigma_2^2
\Big)(\pi_2^a)^2\nonumber\\ &&\!\!\!\! + \lambda_1 \sigma_1^4
 + \lambda_2 \sigma_2^4
+ (\lambda_3 + \lambda_4) \sigma_1^2 \sigma_2^2 + \lambda_5 \sigma_1^3 \sigma_2
+ \lambda_6 \sigma_1
\sigma_2^3 .\nonumber \label{effpot11}
\end{eqnarray}

The current quark mass $m_q$ corresponds to the average of the external scalar sources
$M_j(x) = s_j(x) + i \tau^a p^a_j(x)$, namely, $\langle M_j(x)\rangle = -\frac12 d_j m_q$ and
thus the relevant new terms beyond the chiral limit  can be produced with the help of the
formal replacement $H_j\rightarrow c_jm_q$ in all quadratic and quartic operators
included in \gl{effpot1}  and by adding these new terms with new constants into the effective
potential. We will consider the two flavor case and retain the terms softly breaking chiral symmetry and  linear in $H_j$ and $m_q$,
thereby neglecting  terms cubic in scalar fields exploiting the non-linear equivalence
transformation \[  H_j \rightarrow H_j +\sum_{k,l,m = 1,2} b_{jklm}H_k H^\dagger_l H_m .\]
It corresponds to the choice of external scalar sources linear in $H_j$, $\sum_{j=1,2}\mbox{\rm tr  }(M^\dag_j H_j+H_j^{\dag}M_j)$ . Thus
we add two new  terms to our effective potential \gl{effpot11}
\begin{equation}
-\frac12 m_q
\mbox{\rm tr  }
\Big[d_1(H_1+H_1^{\dag})+ d_2(H_2+H_2^{\dag})\Big].
\end{equation}
Making use of our chiral parametrization of the fields $H_j$ through the chiral field
\begin{equation}
U\equiv \xi^2 = \cos\frac{|\pi_1^a|}{F_0}+i\frac{\tau^a\pi_1^a}{|\pi_1^a|}\sin\frac{|\pi_1^a|}
{F_0},
\end{equation}
one derives the following extension of the effective potential \gl{effpot11}
\begin{eqnarray}
\Delta
V_{\mbox{\rm eff  }}(m_q)&=&
2 m_q\left[-(d_1\sigma_1+d_2\sigma_2)\cos\frac{|\pi_1^a|}{F_0}\right.\nonumber\\ &&\left.+
d_2\frac{\pi_1^a\pi_2^a}{|\pi_1^a|}\sin\frac{|\pi_1^a|}{F_0}\right]. \label{currmass}
\end{eqnarray}
The effective potential \gl{effpot11}, \gl{currmass} will be used to search for CSB and for the
derivation of meson masses.

\subsection{Mass-gap equations and second variations od effective potential}

Let us now investigate the possible appearance of a non-zero v.e.v.'s  of
pseudoscalar fields. Some time ago it was proved in \cite{witten} that parity and vector
flavor symmetry could not undergo spontaneous symmetry breaking in a vector-like
theory such as QCD at normal vacuum conditions at zero
chemical potential . Finite baryon density however may result in a breaking of parity invariance
by simply circumventing the hypothesis of the theorem. Indeed
the presence of a finite chemical potential leads to the appearance of a constant imaginary
zeroth-component of a vector field and the conditions under which the results of \cite{witten}
were proven are not fulfilled anymore.

Accordingly let us check the possibility of condensation of the neutral isospin pseudoscalar components (in order not to violate charge conservation),
\begin{equation}
\pi_1^a=\pi^0\delta^{3a},\qquad
\pi_2^a=\rho\delta^{3a},
\end{equation}
and for the vacuum solutions take $\pi_1^\pm = \pi_2^\pm = 0$.
In this case one obtains four mass-gap equations as the pion condensate $\langle\pi^0\rangle \not=0$ becomes, in
principle, possible, unlike in the chiral limit,
\begin{eqnarray}  &&-2(\Delta_{11} \sigma_1 + \Delta_{12} \sigma_2) -
2m_q d_1\cos\frac{\pi^0}{F_0} + 4
\lambda_1 \sigma_1^3 \nonumber\\ &&+ 3\lambda_5 \sigma_1^2 \sigma_2 + 2 (\lambda_3 +
\lambda_4) \sigma_1
\sigma_2^2  + \lambda_6 \sigma_2^3\nonumber\\ && + \rho^2 \Big(2(\lambda_3 - \lambda_4)
\sigma_1 + \lambda_6
\sigma_2\Big) = 0 ,
\label{mg1}\\
&&-2(\Delta_{12} \sigma_1 +\Delta_{22} \sigma_2) -
2 m_q d_2\cos\frac{\pi^0}{F_0} +
\lambda_5 \sigma_1^3 \nonumber\\ &&+ 2 (\lambda_3 + \lambda_4)  \sigma_1^2 \sigma_2 +
3 \lambda_6  \sigma_1
\sigma_2^2 + 4 \lambda_2 \sigma_2^3 \nonumber\\ && + \rho^2 \Big(\lambda_6  \sigma_1 +4
\lambda_2 \sigma_2 \Big) = 0,
\label{mg2}\\&&(d_1\sigma_1+d_2\sigma_2)\sin\frac{\pi^0}{F_0} + d_2\rho\cos\frac{\pi^0}{F_0} = 0, \label{mg4} \end{eqnarray}
\begin{eqnarray}
&&m_q d_2\sin\frac{\pi^0}{F_0} + \rho\Big( - \Delta_{22}
+(\lambda_3 - \lambda_4)
\sigma_1^2 + \lambda_6 \sigma_1 \sigma_2\nonumber\\ &&+ 2\lambda_2 (\sigma_2^2 + \rho^2)\Big)=0\no &&=\rho\Big(
-\frac{m_qd_2^2}{(d_1\sigma_1+d_2\sigma_2)}\cos\frac{\pi^0}{F_0}- \Delta_{22}
+(\lambda_3 - \lambda_4)
\sigma_1^2\no && + \lambda_6 \sigma_1 \sigma_2+ 2\lambda_2 (\sigma_2^2 +\rho^2)\Big),\label{mg3}
\end{eqnarray}
where the last equality follows from eq.~\gl{mg4}. As well the equality $\rho=0$ entails $\pi^0=0$ from eq.~\gl{mg4} if
$d_1\sigma_1+d_2\sigma_2\neq 0$ and $d_2\neq 0$. However, as will be seen below,
the combination $d_1\sigma_1+d_2\sigma_2$ is related to the quark condensate
\begin{equation} \label{gor}
\langle d_1\sigma_1+d_2\sigma_2\rangle =- \langle\bar{q}q\rangle > 0,
\end{equation}
hence, this combination cannot be zero. For $d_2=0$ one has always $\langle\pi^0\rangle=0$
and the parity breaking pattern remains the same as for the massless
case. We neglect the possibilities $\langle\pi^0\rangle=F_0 n \pi, n = 0,\pm1,\pm2,\ldots,$ as not relevant for the physics studied in this paper. For $d_2 \neq 0$  both pseudoscalar v.e.v. $\langle\pi^0\rangle$ and
$\rho$ can arise simultaneously only. To avoid spontaneous parity breaking in then normal vacuum
of QCD, it is thus {\it sufficient} to impose,
\begin{equation}
(\lambda_3 - \lambda_4) \sigma_1^2
+ \lambda_6 \sigma_1
\sigma_2 + 2\lambda_2 \sigma_2^2  - \Delta_{22} -
\frac{m_qd_2^2}{(d_1\sigma_1+d_2\sigma_2)} > 0,
\label{ineq1}
\end{equation}
on the mass-gap solutions $\sigma_j= \langle\sigma_j\rangle$ in the vicinity of a minimum of effective potential. It follows from the last line in eq.~\gl{mg3}.
Since QCD in normal conditions does not lead to parity
breaking, the low-energy model must fulfill \gl{ineq1}.

For the parity-even vacuum state  the necessary condition to have a minimum for non-zero
$\sigma_j= \langle\sigma_j\rangle$  (for vanishing $\rho$), equivalent to the condition of having CSB in QCD, can be
 derived from the condition to get a local maximum (or at least a saddle point) for zero $\sigma_j$.
At this point the extremum is characterized by the matrix $-\Delta_{jk}$ in (\ref{effpot1}).
It must have at least one negative eigenvalue. This happens either for ${\rm Det} \Delta > 0,\ \tr{\Delta} > 0$
(maximum at the origin) or for ${\rm Det} \Delta < 0$ (saddle point at the origin) at $\sigma_j= \langle\sigma_j\rangle$  . The sufficient
conditions follow from the positivity of the second variation for a non-trivial solution of the two
first equations \gl{mg1}, \gl{mg2} at $\rho = 0$. The matrix containing the second variations
$\hat V^{(2)}$  for the scalar sector is
\begin{eqnarray}
&&\!\!\!\!\frac12 V^{(2)\sigma}_{11} = -\Delta_{11} + 6\lambda_1 \sigma_1^2 + 3\lambda_5
\sigma_1 \sigma_2 +
(\lambda_3 + \lambda_4)  \sigma_2^2 , \nonumber\\ && \!\!\!\!V^{(2)\sigma}_{12} = - 2\Delta_{12}
+ 3 \lambda_5
\sigma_1^2 + 4  (\lambda_3 + \lambda_4) \sigma_1 \sigma_2 + 3\lambda_6
\sigma_2^2  , \nonumber\\ &&\!\!\!\!\frac12
V^{(2)\sigma}_{22} = -  \Delta_{22} +  (\lambda_3 + \lambda_4)  \sigma_1^2 + 3
\lambda_6  \sigma_1
\sigma_2 + 6 \lambda_2 \sigma_2^2 .\  \label{secvar}
\end{eqnarray}
In turn, the nonzero elements of the second variations $\hat V^{(2)}$ in the pseudoscalar sector are
\begin{eqnarray}
&&\!\!(V^{(2)\pi}_{jk})^{ab} \equiv V^{(2)\pi}_{jk} \delta^{ab};\no&&\!\!\frac12
V^{(2)\pi}_{11} =  m_q\frac{d_1\sigma_1+d_2\sigma_2}{F_0^2},
\quad V^{(2)\pi}_{12} = \frac{2 m_q d_2}{F_0}, \nonumber\\ &&\!\!\frac12
V^{(2)\pi}_{22} =
- \Delta_{22} +(\lambda_3 - \lambda_4) \sigma_1^2 + \lambda_6
\sigma_1 \sigma_2 +
2\lambda_2 \sigma_2^2. \label{secvar1}
\end{eqnarray}
The required conditions are given by $\tr{\hat V^{(2)}} >0$ and $\mbox{\rm Det}\hat V^{(2)} >0$ at $\sigma_j= \langle\sigma_j\rangle$ .
For positive matrices it means that
\begin{equation}
V^{(2)\sigma}_{jj} >0;\quad V^{(2)\pi}_{kk} > 0. \label{ineq2}
\end{equation}
The diagonalization of the  matrix $(V^{(2)\pi}_{jk})$ leads to physical mass states for pseudoscalar mesons $\pi, \Pi$ which are the mixtures of $\pi_1, \pi_2$ .
 The eigenvalues of \gl{secvar1} eventually give their masses squared  and
thereby must be positive according to the inequality \gl{ineq1}. The latter corresponds
to the positivity of the determinant,
\ba
&&\mbox{\rm Det}\hat V^{(2)\pi} = 4 m_q\frac{d_1\sigma_1+d_2\sigma_2}{F_0^2}\Big( - \Delta_{22} +(\lambda_3 - \lambda_4) \sigma_1^2\nonumber\\ &&
+ \lambda_6 \sigma_1
\sigma_2 + 2\lambda_2 \sigma_2^2  -
\frac{m_q d_2^2}{(d_1\sigma_1+d_2\sigma_2)}\Big),
\ea wherefrom it becomes evident that the inequality \gl{ineq1} is also a {\it necessary} condition for the absence of spontaneous parity breaking. Indeed it follows from the positivity of matrix element $V^{(2)\pi}_{11}$
 that the combination $\langle d_1\sigma_1+d_2\sigma_2\rangle > 0.$ In fact, to the leading order
in $m_q$ the masses of a lighter $\pi$ and a heavier $\Pi$ mesons are proportional to $V^{(2)\pi}_{11}$ and $V^{(2)\pi}_{22}$,
respectively (see  \gl{secvar1}). A more detailed analysis of the pseudoscalar meson spectrum will be given in Subsect.\ref{pimass}. The requirement to have a positive determinant of the matrix $V^{(2)\pi}_{jk}$ is supported by \gl{ineq1}.

The two set of conditions, namely those presented in eq. (\ref{ineq1}) and in eq. (\ref{ineq2})
represent restrictions that the symmetry breaking pattern of QCD imposes on its
low-energy effective realization at vanishing chemical potential.

One can easily find the correction linear in $m_q$ to the vacuum solution in the
chiral limit
\begin{eqnarray}
&&\langle\sigma_j\rangle(m_q) \simeq \langle\sigma_j\rangle(0) + 2 m_q \Delta^\sigma_j;\no
&&\vec\Delta^\sigma \equiv
\left(\begin{array}{c}\Delta^\sigma_1\\ \Delta^\sigma_2 \end{array}\right)=
\Big(\hat V^{(2)\sigma}\Big)^{-1}\cdot\, \vec d \no
&&=
\frac{1}{\mbox{\rm  Det}\hat V^{(2)\sigma}}\left(\begin{array}{c} d_1
V^{(2)\sigma}_{22} -d_2 V^{(2)\sigma}_{12} \\d_2  V^{(2)\sigma}_{11} - d_1
V^{(2)\sigma}_{12}   \end{array}\right);\quad \vec d \equiv
\left(\begin{array}{c}d_1\\d_2 \end{array}\right) .\nonumber
\end{eqnarray}
Using these equations the corrections to the masses of scalar and heavy
pseudoscalar mesons can be derived straightforwardly. In particular, for scalar mesons the
corrections to the mass matrix are
\begin{eqnarray}
\Delta V^{(2)\sigma}_{kl} &=& 2 m_q \sum_{j,m =1,2}\partial_{j}
V^{(2)\sigma}_{kl} \big(\hat V^{(2)\sigma}\big)^{-1}_{jm} d_m\,
\quad \Big[\partial_j \equiv \partial_{\sigma_j}\Big]\,\no &=& 2 m_q
\sum_{j,m =1,2}\partial_{k} V^{(2)\sigma}_{lj} \big(\hat
V^{(2)\sigma}\big)^{-1}_{jm} d_m \no &=&2 m_q  \partial_{k} \Big(\hat
V^{(2)\sigma}\Big)\cdot \Big(\hat V^{(2)\sigma}\Big)^{-1}\cdot\, \vec d  ,
\end{eqnarray}
whereas in the pseudoscalar sector
\begin{equation}
\Delta V^{(2)\pi}_{22} = 2 m_q \sum_{j,m =1,2}\partial_{j} V^{(2)\pi}_{22}
\big(\hat V^{(2)\sigma}\big)^{-1}_{jm} d_m \, .
\end{equation}
The latter term saturates the current quark mass correction for heavy pseudoscalar meson masses.

\section{Reduction of coupling constants and extrema of effective potential \label{reduc}}
Let us investigate how many extrema the effective potential possesses for different values of the
coupling constants. In this Section we take the chiral limit $m_q = 0$ for simplicity.
It turns out that when the chemical potential and temperature are zero one can eliminate one of the
constant in the effective potential by a redefinition of the fields. Indeed, one can change the variable
\begin{equation}
H_2 =  \alpha H_1 + \beta \tilde H_2 , \label{lintr}
\end{equation}
using a  linear transformation with real coefficients $\alpha, \beta$ (to preserve reality
of $\tilde\sigma_j, \pi_j^a$). With the help of this redefinition one can diagonalize the quadratic
part in \gl{effpot1} and make its coefficients equal
$\tilde\Delta_{11} = \tilde\Delta_{22} =\det\hat\Delta /\Delta_{22}\equiv \Delta$.
Then
\begin{equation}
\sum_{j,k=1}^2
\tr{H^\dagger_j \Delta_{jk} H_k} = \Delta \tr{H^\dagger_1 H_1+ \tilde H^\dagger_2 \tilde H_2} .
\end{equation}
A further reduction of the coupling constants affects the dependence of free energy on
finite chemical potential and temperature (see below), but it can be implemented when both external
control parameters vanish; namely we perform the orthogonal rotation of two fields
\ba
&&H_1 = \cos\phi \check H_1 + \sin\phi \check H_2,\no&& \tilde H_2 = -\sin\phi \check H_1 +
\cos\phi \check H_2  . \label{lintrans}
\ea
Then the coefficient in the operator
$(\check H^\dagger_1 \check H_2 + \check H^\dagger_2 \check H_1) \check H^\dagger_1 \check H_1 $
becomes equal to
\begin{eqnarray}
\frac{\check\lambda_5}{\cos^4\phi} &=& \lambda_5 - 2(\lambda_3 + \lambda_4 -
2\lambda_1) \tan\phi \nonumber\\
 &&- 3 (\lambda_5 - \lambda_6) \tan^2\phi+2(\lambda_3 + \lambda_4 - 2\lambda_2) \tan^3\phi\nonumber\\
 && - \lambda_6 \tan^4\phi \equiv {\cal
P}_{\lambda_5} (\tan\phi) .\label{lambda5}
\end{eqnarray}
One can always fix $\lambda_6 < 0$ by reflection of $\check H_2$ . Then if $\lambda_5 < 0$ then
${\cal P}_{\lambda_5} (0) < 0 $ but evidently for $\tan\phi \gg 1$, ${\cal P}_{\lambda_5}
(\tan\phi)\sim - \lambda_6 \tan^4\phi  > 0$ and therefore the equation ${\cal P}_{\lambda_5}
(\tan\phi) = 0$ has at least one (positive) real root . In the complementary region  $\lambda_5 \geq
0$ and therefore ${\cal P}_{\lambda_5} (0) > 0 $. In this case one can look at $\tan\phi = \pm 1 $
where
\begin{equation}
{\cal P}_{\lambda_5}(\pm 1) = - 2(\lambda_5 -\lambda_6) \pm 4(\lambda_1
-\lambda_2) ,
\end{equation}
so that one of these combinations is negative. Again the comparison with
the asymptotics allows to conclude that there is a real root for ${\cal P}_{\lambda_5} (\tan\phi) =
0$. Thus for any sign of $\lambda_5$ it can be eliminated by a proper rotation of scalar fields.

Let us take the basis of operators with $\check\lambda_5 = 0$ . Then, after renaming the fields
\begin{eqnarray}
V_{\mbox{\rm eff  }}&=& - \Delta \Big((\sigma_1)^2+(\sigma_2)^2\Big)+  \check\lambda_2
\Big((\pi_2^a)^2\Big)^2\nonumber\\ && + (\pi_2^a)^2 \Big(- \Delta +(\check\lambda_3 - \check\lambda_4) \sigma_1^2 + \check\lambda_6
\sigma_1 \sigma_2 + 2\check\lambda_2 \sigma_2^2 \Big)\nonumber\\ && + \check\lambda_1 \sigma_1^4
 + \check\lambda_2 \sigma_2^4
+ (\check\lambda_3 + \check\lambda_4) \sigma_1^2 \sigma_2^2 + \check\lambda_6 \sigma_1 \sigma_2^3 .\label{veff}
\end{eqnarray}
This potential simplifies the mass gap equations and second variations in order to investigate
their solutions analytically. The effective potential must provide the familiar CSB
at normal conditions ($\mu=T=0$). Thus in the chiral limit there are at least
two minima  related by the symmetry rotation $\sigma_{1,2} \rightarrow -\sigma_{1,2}$ and one maximum
at the origin. This is implemented by assigning a real singlet v.e.v. $\langle\sigma_1\rangle > 0$
to $ \check H_1$ thereby selecting one of the minima.

In this Section we shall assume $\check\lambda_5=0$ in order to determine the different vacua of the theory at
zero temperature and chemical potential.

\subsection{Search for the extrema of  effective potential}
In the parity symmetric case the second eq. \gl{mg2} reads
\begin{equation}
\sigma_2 (- \Delta + (\check\lambda_3 + \check\lambda_4)  \sigma_1^2  + \frac32 \check\lambda_6
\sigma_1 \sigma_2 + 2 \check\lambda_2 \sigma_2^2) = 0 . \label{muzero}
\end{equation}
One of its solutions is $\sigma^{(0)}_2 = 0$ and directly from eq.\gl{mg1} one finds
\begin{equation} \sigma^{(0)}_2 = 0,\quad  (\sigma^{(0)}_1)^2 =
\frac{\Delta}{2\check\lambda_1} .
\end{equation}
For stable solutions $\check\lambda_1 >0$ and therefore $\Delta > 0$.

Another set of solutions $\sigma^{(m)}_{1,2};\ m = 1,2,3$ comes from eq. \gl{muzero} for $\sigma_2 \not= 0$ .
With a combination of the mass-gap eqs. \gl{mg1} and \gl{muzero} one can decouple the equation in
terms of the ratio $t = \sigma_2/\sigma_1$,
\begin{eqnarray}
&&{\cal P}_3 (t)= t^3 - a t^2 - b t  +c = 0,\label{cubic}\\
&&a= \frac{2\big(\check\lambda_3 + \check\lambda_4 - 2\check\lambda_2\big)}{-\check\lambda_6},\quad b= 3,\no
&& c=
\frac{2\big(\check\lambda_3 + \check\lambda_4- 2\check\lambda_1\big)}{-\check\lambda_6} , \nonumber
\end{eqnarray}
where the sign is fixed for $\check\lambda_6 < 0$ and $c> 0$ as it is shown in the next subsection.
As the order of the equation is odd there may one or three (as in Fig.1) real solutions.
Because ${\cal P}_3 (0) > 0,\ {\cal P}'_3 (0) < 0$ one concludes that one of the solutions is negative.
\begin{figure}
[tbp]
\begin{minipage}{\columnwidth}
\hspace{-1.5cm}
\includegraphics[width=1.5\textwidth]{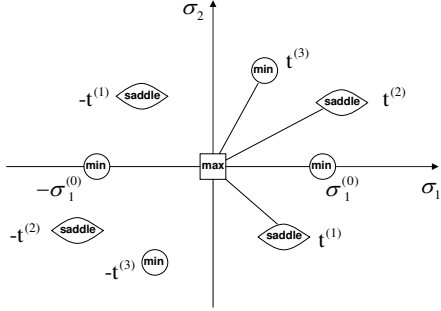}
\end{minipage}

\vspace{-3.0cm}
\caption{\sl Extrema of effective potential in the reduction basis: the maximum
is placed in the square, four minima are located in the circles and the corresponding
four saddle points are depicted by the lentils. The existence of two solutions $t^{(2)}, t^{(3)}$ with positive values
of $\sigma_j$ is governed by condition (\ref{ccc}). Which one corresponds to the true minimum depends on
the actual value of the phenomenological constants.}\nonumber
\end{figure}

Let us analyze the extrema of
${\cal P}_3 (t)$
\ba &&{\cal P}'_3 (t) = 0 \longrightarrow t^2 - \frac23 a t - b = 0,\nonumber\\
&&
t_{\pm} = \frac13 a \pm \sqrt{\frac19 a^2 +b},\ t_+ > 0,\ t_- < 0 .
\ea
All together it means that a negative solution $t^{(1)} < 0$ always exists and (if any) two
more solutions are positive, $t^{(2)} < t^{(3)}$,
and separated by a minimum of cubic polynomial ${\cal P}_3$ . Therefore the existence of two positive solutions is regulated by the sign of
${\cal P}_3 (t_+)$. They exist if
\ba
{\cal P}_3 (t_+)&& =  c -
\frac{a}{3} \left(\frac{a}{3}+ \sqrt{\frac{a^2}{9} + b}\right)^2 < 0; \no c && < a \dfrac{\frac{a}{3}+ \sqrt{\frac{a^2}{9} + b}}{-\frac{a}{3}+ \sqrt{\frac{a^2}{9} + b}}.\label{ccc}
\ea
Evidently it takes place for some positive $a$, i.e. for
\begin{equation}
\check\lambda_3 +\check\lambda_4 > 2\check\lambda_2 .
\end{equation}

Finally we can have at most two minima for positive $\sigma_1$, namely, at  $\sigma^{(0)}_2 = 0$ and at $t^{(2)}$ or $t^{(3)}$ which entails two more minima for negative
$\sigma_1$ due to symmetry under $\sigma_j \rightarrow -\sigma_j$. Then  other solutions correspond to saddle points as four minima must be separated by  four saddle
points situated around the maximum, see Fig.1. This configuration is unique for potentials bounded below, namely, any saddle point connects two adjacent minima. Later on we will see that in order to implement a first-order phase transition to  stable
nuclear matter we just need two minima for positive $\sigma_1$ . Thereby, in the half plane of positive $\sigma_1$  one has to reveal four
solutions, namely, one is $\sigma^{(0)}_2 = 0$ and three for $\sigma_2 \not= 0$ which are inevitably mark one more minimum and two saddle points. Thereby the condition (\ref{ccc})
should be satisfied in order to be able to describe the saturation point .

After the appropriate roots $t_j$ are found one can use Eq. \gl{muzero} and find the v.e.v. of $\sigma_j$,
\be
\sigma_1^2 = \frac{\Delta}{(\check\lambda_3 + \check\lambda_4)  + \frac32 \check\lambda_6
t_j + 2 \check\lambda_2 t^2_j} >0 ;\quad \sigma_2 = t_j \sigma_1. \label{sigma1}
\ee
The latter inequality holds for any $t_j$ if
\be
\check\lambda_2 (\check\lambda_3 + \check\lambda_4) \geq \frac{9}{32} (\check\lambda_6)^2 .
\ee
Otherwise the existence of real $\sigma_1$ for solutions  $t_j$ needs a more subtle investigation.

Let us recall that all the inequalities obtained in this Section are referred to the field basis with
fully diagonal $\Delta_{ij} = \Delta \delta_{ij}$ and with $\check\lambda_5 = 0$. However it is evident that the qualitative structure of extrema is independent of the basis choice.
\subsection{Selection of the minima}
In all cases the conditions of minimum come from the positive definiteness of the matrix of second
variations of effective potential,
\begin{eqnarray}
\frac12 V^{(2)\sigma}_{11} &=& -\Delta + 6\check\lambda_1 \sigma_1^2
+ (\check\lambda_3 + \check\lambda_4)  \sigma_2^2 > 0,  \label{secvar2}\\ V^{(2)\sigma}_{12} &=&  4  (\check\lambda_3 +
\check\lambda_4) \sigma_1 \sigma_2 + 3\check\lambda_6 \sigma_2^2  , \nonumber\\ \frac12 V^{(2)\sigma}_{22}
&=& -  \Delta +  (\check\lambda_3 + \check\lambda_4)  \sigma_1^2 + 3 \check\lambda_6  \sigma_1 \sigma_2 + 6 \check\lambda_2
\sigma_2^2 > 0,\nonumber\\ \frac12 V^{(2)\pi}_{22}&=& -\Delta+(\check\lambda_3 - \check\lambda_4) \sigma_1^2
+ \check\lambda_6 \sigma_1\sigma_2 + 2\check\lambda_2 \sigma_2^2 > 0;\nonumber\\ V^{(2)\pi}_{11}&=& V^{(2)\pi}_{12} = 0 . \nonumber
\end{eqnarray}
For $\sigma_2^{(0)} = 0$ they  read
\begin{equation} (
\check\lambda_3 \pm \check\lambda_4)  > 2  \check\lambda_1, \longrightarrow \check\lambda_3 > |\check\lambda_4| \label{ineq12}
\end{equation}
for $\Delta > 0$ as it is required by the absence of chiral collapse (and the spectrum at the SPB
point, see below). It gives support to the condition $c>0$ in the previous subsection.

For $\sigma_2^{(m)} \not= 0$ one obtains a number of bounds on the solution from the second variation
\begin{eqnarray}
\frac12 V^{(2)\sigma}_{11} &=&  (\sigma_1)^2 \Big[4\check\lambda_1 - \frac12 \check\lambda_6
t^3\Big]  > 0, \nonumber\\ V^{(2)\sigma}_{12} &=& (\sigma_1)^2 \Big[ 4  (\check\lambda_3 +
\check\lambda_4) t + 3\check\lambda_6 t^2 \Big]  , \nonumber\\ \frac12
V^{(2)\sigma}_{22} &=& (\sigma_1)^2 \Big[\frac32 \check\lambda_6  t + 4 \check\lambda_2 t^2 \Big] >
0,\nonumber\\ \frac12 V^{(2)\pi}_{22}&=& (\sigma_1)^2 \Big[ - 2\check\lambda_4  - \frac12 \check\lambda_6 t\Big]  > 0.
\label{secvar3}
\end{eqnarray}
Evidently if $\check\lambda_4 > 0$ then $\sigma_2 > 0 \rightarrow t > 0$.
The remaining bound must come from the positivity, $ \det\hat V^{(2)} > 0$.

\section{\label{finite-mu} Finite temperature and baryon chemical
potential}
\subsection{Coupling the effective lagrangian to the environment}
We are building a model of meson medium starting from the quark sector of QCD.
Its thermodynamical properties and relationship to a dense baryon matter will be examined with the
help of thermodynamical potentials derived from the constituent quark model in the large $N_c$ (mean field)
approach. This gives a prescription to connect the properties of quark and nuclear matter and estimate the
parameters of our model to reproduce meson phenomenology and the bulk characteristics of nuclear matter
such as binding energy, normal nuclear density and (in)compressibility.

The meson degrees of freedom  present in our model appear after bosonization of QCD in the vacuum and the relevant effective potential is given in Sec. \ref{sigma-model}, eq.~\gl{effpot1}.
The effects of infinite homogeneous baryon matter on the effective meson lagrangian are described by the
baryon chemical potential $\mu$, which is transmitted to the meson lagrangian  via a local quark-meson
coupling (in the leading order of chiral expansion $\mu^2/\Lambda^2$). In turn, in the large $N_c$ limit
one can neglect the temperature dependence due to meson collisions. The temperature
$T$ is induced with the help of the imaginary time Matsubara formalism for quark Green functions\cite{matsu}
\begin{equation}
\omega_n = \frac{(2n+1)\pi}{\beta} ,\quad \beta = \frac{1}{kT} .
\end{equation}
For real physics with 3 colors this approximation to thermal properties of
mesons is expected to be less precise as meson loops contribute substantially
to the thermodynamic characteristics for large temperatures (first of all a hot pion gas). Nevertheless it should be sufficient to describe
qualitatively the interplay between baryon density and temperature at the phase
transition.

Without loss of generality we can specify one of the collective fields $H_j$, namely, $H_1$ as that one which
has local coupling to quarks: this actually defines the chiral multiplet $H_1$. The set of
coupling constants in \gl{effpot1} is sufficient to support this choice as well as to fix the Yukawa
coupling constant to unity. Accordingly, we select the basis in which finite density and temperature were transmitted to
the boson sector by means of
\begin{equation}
\Delta {\cal
L}_q=  \bar q_R H_1 q_L +\bar
q_L H_1^\dagger q_R \longrightarrow  \bar Q \sigma_1 Q,
\end{equation}
where $Q_L = \xi q_L, Q_R =\xi^\dagger q_R; \xi = \exp\{i\hat\pi/2F_0\}$ stand for constituent quarks \cite{aet}. Then for finite temperatures and chemical potentials the free Fermi gas contribution to the generalized $\sigma$ model lagrangian originates from the quark action in Euclidean space-time (thermal field theory) \cite{kapusta},
\ba
{\cal S}_q &=& \int\limits_0^\beta d\tau\int d^3 x \ q^\dagger\Big( \not\!\partial -\gamma_0 \mu + ( H_1 P_L + H_1^\dagger P_R) \Big) q\no
&\simeq& \int\limits_0^\beta d\tau\int d^3 x \bar Q(\not\!\partial -\gamma_0 \mu + \sigma_1 ) Q,
\ea
where $P_{L,R} \equiv \frac12 (1\pm \gamma_5) $. As we want to calculate the effective potential we neglect the gradient of chiral fields $\partial_\mu \xi \sim 0$ in the last expression.

After averaging over constituent quarks one obtains the free Fermi gas contribution to the vacuum effective potential \cite{kapusta}
\ba
&&\Delta V_{\mbox{\rm eff  }}(\sigma_1, \mu, \beta) = V_{\mbox{\rm eff},Q} (\sigma_1, \mu, \beta)- V_{\mbox{\rm eff},Q} (\sigma_1, 0, 0)\no&=& -\frac{{\cal
N}}{\beta} \int \frac{d^3p}{\pi }\, \Big\{ \log\Big(1+\exp(-\beta(E-\mu))\Big)\no&& + \log\Big(1+\exp(-\beta(E+\mu))\Big)\Big\}\no&=&-\frac{4{\cal
N}}{\beta} \int\limits_{\sigma_1}^{\infty} dE E \sqrt{E^2-\sigma_1^2} \Big\{ \log\Big(1+\exp(-\beta(E-\mu))\Big)\no&& + \log\Big(1+\exp(-\beta(E+\mu))\Big)\Big\} \label{deltaveff}\\
 &=&
- \frac43 {\cal N}\int\limits_{\sigma_1}^{\infty}\, dE \Bigl(E^2-\sigma_1^2\Bigr)^{3/2} \,
\frac{\cosh(\beta\mu) + \exp(-\beta E)}{\cosh(\beta\mu) + \cosh(\beta E)}, \nonumber
\ea
where $E = \sqrt{p^2 +\sigma_1^2}$  .
The last expression for the Fermi gas free energy in \gl{deltaveff} can be obtained with the help of integration by part and an appropriate change of energy and momentum variables.

Accordingly the complete effective potential is specified as,
\be
 V_{\mbox{\rm eff  }}(\sigma_j, \pi_j^a; \mu, \beta) = V_{\mbox{\rm eff  }}(\sigma_j, \pi_j^a; 0, 0) + \Delta V_{\mbox{\rm eff  }}(\sigma_1, \mu, \beta) . \nonumber
\ee
Following this recipe the quark temperature and chemical potential dependence can be
derived for mass gap equations --  the conditions for a minimum of the
effective potential. Namely, taking into account the choice of variables \gl{chipar} the first equation
\gl{mg1} is modified to
\begin{eqnarray}
&&-2\Delta
\sigma_1  - 2 m_q
d_1\cos\frac{\pi^0}{F_0} + 4 \lambda_1 \sigma_1^3 + 3\lambda_5
\sigma_1^2 \sigma_2 \nonumber\\ &&+ 2 (\lambda_3 + \lambda_4) \sigma_1 \sigma_2^2  + \lambda_6
 \sigma_2^3  + \rho^2 \Big(2(\lambda_3 - \lambda_4) \sigma_1 + \lambda_6
\sigma_2\Big) \nonumber\\ &&+ 2{\cal N}\sigma_1 {\cal A}(\sigma_1, \mu, \beta) = 0,\label{var1}
\end{eqnarray}
with notation ${\cal N} \equiv \frac{N_{c}N_f}{4\pi^2} $. In turn
\begin{eqnarray}
&&{\cal A}(\sigma_1, \mu, \beta)   = \frac{1}{2{\cal N}\sigma_1}\partial_{\sigma_1} \Delta V_{\text eff} (\sigma_1, \beta, \mu) \nonumber\\ &&= 2 \eint \, \sqrt{E^2-\sigma_1^2} \, \frac{\cosh(\beta\mu) + \exp(-\beta
E)}{\cosh(\beta\mu) + \cosh(\beta E)}.
\label{Aend}
\end{eqnarray}
When using the gap equations \gl{mg2}, \gl{mg4}, \gl{mg3}and \gl{var1} one finds the value of the effective potential
at its minima ("on shell"),
\begin{eqnarray}
&&V_{\mbox{\rm eff  }}(\sigma_j, \pi_j^a; \mu, \beta)\left|_{\sigma_j = \langle\sigma_j\rangle;\ \pi_1^a = \delta^{3a}\langle\pi^0\rangle;\ \pi_2^a = \delta^{3a}\rho}\right.\no&&\equiv \widetilde V_{\mbox{\rm eff}}(\mu, \beta) = - \frac12 \Delta \Big(\langle\sigma_1\rangle^2 +
\langle\sigma_2\rangle^2 +
\rho^2\Big)\no&& + \frac32 m_q
\left[-(d_1\langle\sigma_1\rangle+d_2\langle\sigma_2\rangle)\cos\frac{\langle\pi^0\rangle}{F_0}+
d_2\rho\sin\frac{\langle\pi^0\rangle}{F_0}\right]\nonumber\\
&&+\Delta V_{\mbox{\rm eff
}}(\langle\sigma_1\rangle, \mu, \beta) - \frac12 {\cal N}\langle\sigma_1\rangle^2 {\cal
A}(\langle\sigma_1\rangle, \mu, \beta);\label{effmubeta}\\&&\Delta V_{\mbox{\rm eff
}} - \frac12 {\cal N}\langle\sigma_1\rangle^2 {\cal A}\no&& = - \frac13 {\cal
N}\!\!\int\limits_{\langle\sigma_1\rangle}^{\infty} dE (4E^2 - \langle\sigma_1\rangle^2)
\Bigl(E^2-\langle\sigma_1\rangle^2\Bigr)^{1/2} \nonumber\\
&&
\times \frac{\cosh(\beta\mu) + \exp(-\beta E)}{\cosh(\beta\mu) + \cosh(\beta E)}.
\label{effmu1}
\end{eqnarray}
In order to derive it we split the vacuum potential \gl{effpot11}, \gl{currmass} into three pieces according to their field dimension,
\be
V_{\mbox{\rm eff  }}(\sigma_j, \pi_j^a; 0, 0)  = V^{(2)}_{\mbox{\rm eff  }} + V^{(4)}_{\mbox{\rm eff  }}+ \Delta V^{(1)}_{\mbox{\rm eff  }}(m_q) . \label{toteff1}
\ee
Next let us multiply eq.~\gl{var1} by $\sigma_1$,  eq.~\gl{mg2} by $\sigma_2$ and eq.~\gl{mg3} by $\rho$ and sum up. In this way the following on-shell identity is obtained,
\ba
&&V^{(4)}_{\mbox{\rm eff  }} (\langle\sigma_j\rangle) = -\frac12 V^{(2)}_{\mbox{\rm eff  }}(\langle\sigma_j\rangle) - \frac14 \Delta V^{(1)}_{\mbox{\rm eff  }}(m_q) (\langle\sigma_j\rangle)\no&& -  \frac12 {\cal N}\langle\sigma_1\rangle^2  {\cal
A}(\langle\sigma_1\rangle, \mu, \beta) .\label{effidentity}
\ea
The final result \gl{effmubeta} can be derived by insertion of this identity in eq. \gl{toteff1}.

Let us notice that the chosen specification of collective fields $H_j$ is compatible with the transformation \gl{lintr} and
therefore one can proceed to the diagonal quadratic part of the potential \gl{effpot1}. However the additional linear
transformation \gl{lintrans}  would split the constituent mass in the quark Yukawa vertex into two fields
$$\bar q_R H_1 q_L + \mbox{\rm  h. c.}\longrightarrow \bar
q_R (\cos\phi \check H_1 + \sin\phi \check H_2) q_L + \mbox{\rm  h. c.}$$ It means that a possible change of
the basis used in Sec. 4 to eliminate the constant $\lambda_5$ would affect the chemical potential driver
\ba&& \Delta V_{\text eff} (\sigma_1, \beta, \mu)\no
\rightarrow&& \Delta V_{\text eff}(\sqrt{(\cos\phi\ \check\sigma_1 + \sin\phi\ \check\sigma_2)^2 +\check\rho^2},
\ \beta, \mu). \ea
Thereby all the  mass gap equations \gl{mg1}--\gl{mg3} would obtain new contributions depending on $T$ and $\mu$ making the equations less tractable. In order to keep the simplified form of the  mass gap equations we prefer to retain the single scalar field in the Yukawa vertex and  include the dependence on environment conditions in one mass-gap equation only . Correspondingly we take, in general, $\lambda_5\not=0$.

 However, the qualitative results derived in the
previous Section on the different vacua for vanishing temperature and chemical potential remain obviously valid. Namely, in the CSB regime one has at most one maximum, four minima and four saddle points at our disposal in order to simulate nuclear matter properties.

\subsection{Thermodynamic properties of the model at $T\neq 0$}
Thermodynamically the system is described by the pressure $P$, the energy density,
$\varepsilon$ and the entropy density $S$. The pressure is determined by the potential density difference with
and without the presence of chemical potential, $dP = -dV$
\begin{equation}
P(\mu,\beta) \equiv \widetilde V_{\mbox{\rm eff  }}(0,0)-
\widetilde V_{\mbox{\rm eff  }}( \mu,\beta), \label{press}
\end{equation}
The energy density is related
to the pressure, baryon density and entropy density by
\begin{equation}
\varepsilon = - P +N_c \mu \varrho_B + T S. \label{enden}
\end{equation}
The chemical potential is defined as
\begin{equation}
\partial_{\varrho_B} \varepsilon = N_c \mu , \label{chem}
\end{equation}
with the entropy and volume held fixed. The factor $N_c$ is introduced to relate
the quark and baryon chemical potentials. Since $\varepsilon$ is independent of $\mu$
\ba
&&\frac{1}{N_c}\partial_\mu P = \varrho_B = -\frac{1}{N_c} \partial_\mu  \widetilde V_{\mbox{\rm eff  }}\label{varrho}\\&& =
\frac{N_f}{\pi^2}\int\limits_{\langle\sigma_1\rangle}^{\infty} dE  E \sqrt{E^2-\langle\sigma_1\rangle^2}
\frac{\sinh(\beta\mu)}{\cosh(\beta\mu) + \cosh(\beta E)} , \nonumber
\ea
where $\langle\sigma_1\rangle = \sigma_1(\mu,\beta)$ on shell.

In turn the entropy is defined as
\begin{equation}
 S= \partial_T P = - \partial_T \widetilde V_{\mbox{\rm eff  }},\quad T S = \beta \partial_\beta \widetilde V_{\mbox{\rm eff}}.
\end{equation}
with the baryon density and volume held fixed.

The above equation allows to calculate the energy density \gl{enden} in our model in terms of the effective potential on shell,
\be
\varepsilon = (- 1 +\mu \partial_\mu -\beta \partial_\beta)\tilde V_{\mbox{\rm eff  }}( \mu, \beta) . \label{enden+}
\ee
\subsection{Zero temperature and finite density}
In this subsection we consider the zero-temperature case  and study the regime of chemical
potentials comparable with the v.e.v. $\sigma_1$. At zero temperature $T=0$ the
contribution from $\mu$ to the effective potential is
 \begin{eqnarray}
&&\Delta V_{\mbox{\rm eff  }}(\sigma_1, \mu) =- \theta(\mu-\sigma_1)\frac43 {\cal N}\int\limits_{\sigma_1}^{\mu}\, dE \Bigl(E^2-\sigma_1^2\Bigr)^{3/2}\nonumber\\
&&=
\frac{{\cal N}}{2}
\theta(\mu-\sigma_1)\Biggl[\mu\sigma_1^2\sqrt{\mu^2-\sigma_1^2} -
\frac{2\mu}{3}(\mu^2-\sigma_1^2)^{3/2}\nonumber\\
&&-
\sigma_1^4\ln{\frac{\mu+\sqrt{\mu^2-\sigma_1^2}}{\sigma_1}}\Biggr]
.\label{potmu}
\end{eqnarray}
The total value of the effective potential at its minimum is
\begin{eqnarray}&&\widetilde V_{\mbox{\rm eff}}(\mu) = - \frac12 \Delta \Big(\langle\sigma_1\rangle^2 +
\langle\sigma_2\rangle^2 +
\rho^2\Big)\nonumber\\
&& + \frac32 m_q
\left[-(d_1\langle\sigma_1\rangle+d_2\langle\sigma_2\rangle)\cos\frac{\langle\pi^0\rangle}{F_0}+
d_2\rho\sin\frac{\langle\pi^0\rangle}{F_0}\right] \no&& - \frac{{\cal N}}{3} \mu \Big(\mu^2 - \langle\sigma_1\rangle\Big)^{3/2} \theta\Big(\mu -\langle\sigma_1\rangle\Big) .\label{effmu}
\end{eqnarray}
Higher-order terms of the chiral expansion in $ 1/\Lambda^2$  are not
considered.

Accordingly in the first mass gap equation \gl{var1}
\begin{eqnarray}
&&{\cal A}(\sigma_1, \mu, \beta) \stackrel{\beta \rightarrow \infty}{=} 2
\theta(\mu - \sigma_1) \int^\mu_{\sigma_1} dE \, \sqrt{E^2-\sigma_1^2}\no&& =\theta(\mu -
\sigma_1)\Big[ \mu
\sqrt{\mu^2-\sigma_1^2}-
\sigma_1^2\ln{\frac{\mu+\sqrt{\mu^2-\sigma_1^2}}{\sigma_1}}\Big] . \label{Amu}
\end{eqnarray}
Then the second variation of effective potential is modified in the only element
\begin{eqnarray}
&&\!\!\!\frac12
V^{(2)\sigma}_{11} = -\Delta_{11}  + 6\lambda_1 \sigma_1^2 + 3\lambda_5
\sigma_1 \sigma_2 + (\lambda_3 +
\lambda_4)  \sigma_2^2 +\nonumber\\ && \!\!\! {\cal N}\theta(\mu-\sigma_1)\left[\mu
\sqrt{\mu^2-\sigma_1^2}-
3\sigma_1^2\ln{\frac{\mu+\sqrt{\mu^2-\sigma_1^2}}{\sigma_1}}\right].
\label{svar1}
\end{eqnarray}
The effective potential \gl{potmu},\gl{effmu} is normalized to reproduce the baryon density
for quark matter
\begin{eqnarray}
&&\varrho_B = -\frac{1}{N_c} \partial_\mu \Delta
V_{\mbox{\rm eff  }}(\sigma_1, \mu)\Big|_{\sigma_1 = \langle\sigma_1\rangle=\sigma_1(\mu)}\label{rhomu}\\ &&= -\frac{1}{N_c} \frac{\mbox{d}
V_{\mbox{\rm eff}}( \mu)}{\mbox{d}\mu} =
\frac{N_f}{3\pi^2} p_F^3 = \frac{N_f}{3\pi^2} \Big(\mu^2-\sigma_1^2(\mu)\Big)^{3/2},
\nonumber
\end{eqnarray}
where the quark Fermi momentum is\\ $p_F = \sqrt{\mu^2-\sigma_1^2(\mu)}$.

\section{Spontaneous parity breaking phase}
\subsection{Mass gap and critical lines for the SPB transition}
Let us examine the possible existence of a critical point, {\it in the chiral limit}
$m_q=0$ for simplicity, where the strict inequality \gl{ineq1} does not hold and
instead for $\mu \geq \mu_{crit}$
\begin{equation}
(\lambda_3 - \lambda_4)
\sigma_1^2 +  \lambda_6 \sigma_1 \sigma_2 +2\lambda_2 \Big(\sigma_2^2    +
\rho^2\Big) = \Delta , \label{creq1}
\end{equation}
so that Eq.\eqref{mg3} admits non-zero values of $\rho$ and thereby SPB arises.
After substituting $\Delta$ from \gl{creq1} into the second eq. \gl{mg1} one finds that
\begin{equation}
  \lambda_5 \sigma_1^2 + 4 \lambda_4 \sigma_1 \sigma_2+
\lambda_6 \Big(\sigma_2^2    + \rho^2\Big)  = 0,\label{creq2}
\end{equation}
where we have taken into account that $\langle\sigma_1\rangle\neq 0$. This, together with \gl{creq1}
completely fixes the v.e.v.'s of the scalar fields $ \sigma_{1,2}$. If $\lambda_2 = 0$ and/or
$ \lambda_6=0$ equations \gl{creq1} or \gl{creq2} unambiguously determine the relation
between $\langle\sigma_1\rangle$ and $\langle\sigma_2\rangle$. Otherwise if  $\lambda_2 \lambda_6 \not=0$ these two
equations still allow to get rid of the v.e.v. of pseudoscalar field leading to the relation
 \begin{eqnarray}
&&\Big(2\lambda_5 \lambda_2 + \lambda_6(\lambda_4 - \lambda_3)\Big) \sigma_1^2
+\Big(8 \lambda_2 \lambda_4 - \lambda_6^2\Big)\sigma_1 \sigma_2 \no&&=  -  \lambda_6 \Delta,
\end{eqnarray}
whose solution is
\begin{eqnarray}  &&\langle\sigma_2\rangle =
A\langle\sigma_1\rangle + \frac{B}{\langle\sigma_1\rangle} > 0;\no&& A \equiv \frac{2\lambda_5 \lambda_2 +
\lambda_6(\lambda_4 - \lambda_3)}{\lambda_6^2 - 8 \lambda_2 \lambda_4 };\quad B
\equiv \frac{\lambda_6 \Delta}{\lambda_6^2 -8
\lambda_2 \lambda_4} . \label{pbrrel}
\end{eqnarray}
Thus in the parity breaking phase the relation between the two scalar v.e.v's is completely
determined and, in particular, does not depend neither on $\rho$ nor on $\mu$.

The first mass gap equation \gl{var1} can be brought to the form
\begin{eqnarray}
&&\Delta  = 2 \lambda_1 \sigma_1^2 + \lambda_5
\sigma_1 \sigma_2 + (\lambda_3 - \lambda_4) (\sigma_2^2  + \rho^2 )\no&&+ {\cal N}{\cal A}(\sigma_1, \mu, \beta),\label{var11}
\end{eqnarray}
if one employs eq. \gl{creq2}. Together with eq. \gl{pbrrel} it allows to find all v.e.v.'s of the
scalar fields $\sigma_j, \rho$ as functions of temperature and chemical potential.

Let us now find the critical value of the chemical potential, namely the
value where $\rho (\mu_c) = 0$, but equations \gl{creq1}, \gl{creq2}, \gl{pbrrel} hold.
Combining the two equations \gl{creq1}, \gl{creq2}
\begin{eqnarray}
\lambda_6  r^2
 + 4\lambda_4  r + \lambda_5  = 0;\quad r \equiv
\frac{\langle\sigma_2\rangle}{\langle\sigma_1\rangle}. \label{homeq}
\end{eqnarray}
In order for a SPB phase to exist this equation has to possess real solutions. If
$ \lambda_6= 0$ there is only one solution corresponding to a
second order transition, but there may exist other solutions that fall beyond
the accuracy of our low energy model (which becomes inappropriate for small values of $\sigma_1$).

We stress that  equations \gl{pbrrel} and  \gl{homeq} contain only the structural
constants of the potential and do not depend on temperature or chemical potential manifestly.
Thus using the critical values
\be
r_{crit} = r_\pm
= \dfrac{-2\lambda_4 \pm \sqrt{4\lambda_4^2 - \lambda_5\lambda_6}}{\lambda_6}
\ee
one can immediately calculate
\ba
&&\langle\sigma_1\rangle^\pm (\Delta, \lambda_j) =\sqrt{\frac{B}{r_\pm - A}};\no&&
\langle\sigma_2\rangle^\pm (\Delta, \lambda_j) = r_\pm \langle\sigma_1\rangle^\pm ,
\ea
where $\langle\sigma_i\rangle^\pm$ are the corresponding critical values.

After substituting these values into equation \gl{var11} for each critical
set of $\langle\sigma_i\rangle$ one derives the boundary of the parity breaking phase
\begin{eqnarray}
&&{\cal N} {\cal A}(\sigma_1^\pm, \mu_{crit}, \beta_{crit}) =
\Delta  - 2\lambda_1 (\langle\sigma_1\rangle^\pm)^2\no&& - \lambda_5 \langle\sigma_1\rangle^\pm \langle\sigma_2\rangle^\pm
- (\lambda_3 - \lambda_4) (\langle\sigma_2\rangle^\pm)^2 . \label{strip}
\end{eqnarray}
It must be positive at critical values of $\langle\sigma_i\rangle^\pm$. The relation \gl{strip} defines a
strip in the $T - \mu$ plane where parity is spontaneously broken. From \gl{Aend} one can obtain
that ${\cal A} >0$ and ${\cal A} \rightarrow \infty$ when $T,\mu \rightarrow \infty$.
It means that for {\it any} nontrivial solution $\langle\sigma_1\rangle^\pm, \langle\sigma_2\rangle^\pm$  the parity breaking
phase boundary exists.

Thus we have proved that if the phenomenon of parity breaking is realized for zero
temperature it will take place in a strip
including lower chemical potentials but higher temperatures.

\subsection{Mass-gap equations in SPB beyond the chiral limit}
Let us now examine again the possible existence of a critical point where
the strict inequality \gl{ineq1} does not hold and for $\mu > \mu_{crit}$
\begin{eqnarray}
&&(\lambda_3 -
\lambda_4) \sigma_1^2 +  \lambda_6 \sigma_1 \sigma_2 +2\lambda_2 \Big(\sigma_2^2
   + \rho^2\Big) -
\Delta\label{creq1m}\\ &&=
\frac{m_q
d_2^2}{(d_1\sigma_1+d_2\sigma_2)}\cos\frac{\pi^0}{F_0}=
\frac{m_q d_2^2}{\sqrt{d_2^2 \rho^2 + (d_1\sigma_1+d_2\sigma_2)^2}}, \nonumber
\end{eqnarray}
where the following consequence of equation\gl{mg4} has been used:
\begin{equation}
\cos\frac{\pi^0}{F_0} =
\frac{d_1\sigma_1+d_2\sigma_2}{\sqrt{d_2^2 \rho^2 +
(d_1\sigma_1+d_2\sigma_2)^2}}. \label{vevpi}
\end{equation}

When combining equation \gl{creq1m} with \gl{mg1}, \gl{mg2} one finds that
\begin{eqnarray}
&&d_1\Big(\lambda_5 \sigma_1^2 + 4 \lambda_4 \sigma_1 \sigma_2+
\lambda_6 (\sigma_2^2    + \rho^2)\Big)\nonumber\\ &&= 2d_2\Bigl(-\Delta+
2\lambda_1 \sigma_1^2 + \lambda_5 \sigma_1 \sigma_2\nonumber\\ && \,\,\,\,+
(\lambda_3-\lambda_4) (\sigma_2^2  + \rho^2)+
 {\cal N}{\cal A}(\sigma_1, \mu, \beta)\Bigr)
,\label{creq2m}\\&&
d_2\Big(\lambda_5 \sigma_1^2 + 4 \lambda_4 \sigma_1 \sigma_2+
\lambda_6 (\sigma_2^2    + \rho^2)\Big) \nonumber\\ && = 2d_1\Big(-\Delta+
(\lambda_3-\lambda_4) \sigma_1^2 \nonumber\\ &&\,\,\,\,+ \lambda_6 \sigma_1 \sigma_2+
2\lambda_2 (\sigma_2^2    + \rho^2)\Big) ,\label{creq3m}
\end{eqnarray}
where we have taken into account that $\langle\sigma_1\rangle\neq 0$. These two relations determine the v.e.v.'s of the
scalar fields $ \sigma_{1,2}$. If $\lambda_2 = \lambda_6=0$ and/or $\lambda_3=\lambda_4, \lambda_6=0$ equations
\gl{creq2m} and \gl{creq3m} firmly fix the relation between $\langle\sigma_1\rangle$ and $\langle\sigma_2\rangle$.
Otherwise an appropriate combination of these two equations still allows us to get rid of the v.e.v.
of the pseudoscalar field\footnote{We recall that in the presence of SPB the distinction between scalars and
pseudoscalars is a nominal one.}. Thus in the parity breaking phase the relation
between the two scalar v.e.v's is completely
determined and in particular does not depend neither on $\rho$ nor on $\mu$.
Using equations \gl{creq1m}, \gl{creq2m} and \gl{creq3m} one can easily eliminate the
variables $\rho$ and $\sigma_2$, obtaining an equation for the variable $\sigma^2_1/\mu^2$ . The
latter completes the determination of the v.e.v.'s.

We notice that in the chiral limit $m_q \rightarrow 0$ the constants $d_{1}, d_2$ become arbitrary
and therefore \gl{creq2m}, \gl{creq3m}  entail three independent relations coinciding with
\gl{creq1}, \gl{creq2}, \gl{var11}.

\section{Approaching the SPB phase transition}
Let us find the character of the phase transition to the SPB phase. In this Section, for brevity, we employ the v.e.v.'s of variables $\sigma_j =\langle\sigma_j\rangle = \sigma_j(\mu), \rho =\langle\rho\rangle= \rho(\mu)$ as functions of the chemical potential $\mu$ on shell. For small values of $\mu-\sigma_1(\mu)>0$,
we know that the value of the odd parity condensate $\langle\rho\rangle$  is zero. Setting $\rho=0$ in
equations \gl{mg1}, \gl{mg2}, \gl{var1} and using \gl{var11} and differentiating w.r.t.
$\mu$ we get
\begin{equation}
\sum\limits_{k=1,2}\hat V^{(2)\sigma}_{jk}\partial_\mu\sigma_k
= - 4 {\cal N}
\sigma_1\sqrt{\mu^2-\sigma_1^2} \delta_{j1} ,
\end{equation}
or, after inversion of the matrix of second
variations,
\begin{eqnarray}
&&\partial_\mu\sigma_1 = - 4 {\cal N}
\sigma_1\sqrt{\mu^2-\sigma_1^2}
\frac{V^{(2)\sigma}_{22}}{{\rm Det}\hat V^{(2)\sigma}} <
0,\no &&\partial_\mu\sigma_2 = 4 {\cal N}
\sigma_1\sqrt{\mu^2-\sigma_1^2} \frac{V^{(2)\sigma}_{12}}{{\rm Det}\hat
V^{(2)\sigma}}. \label{eqmu}
\end{eqnarray}
The possibility of SPB is controlled by the inequality \gl{ineq1}; in order to
approach a SPB phase transition when the chemical potential is increasing we have to diminish the
l.h.s. of inequality \gl{ineq1} and therefore we need to have
 \begin{equation}
\partial_\mu\Big[(\lambda_3 - \lambda_4) \sigma_1^2
+ \lambda_6 \sigma_1 \sigma_2 + 2\lambda_2 \sigma_2^2 -
\frac{m_qd_2^2}{(d_1\sigma_1+d_2\sigma_2)}\Big] < 0.
\end{equation}
This is equivalent (using (\ref{eqmu})) to
\begin{eqnarray}
&&\Big(\lambda_6 \sigma_1 + 4\lambda_2\sigma_2 +
\frac{m_qd_2^3}{(d_1\sigma_1+d_2\sigma_2)^2}\Big)
V^{(2)\sigma}_{12} <\no&&< \Big(2 (\lambda_3 - \lambda_4) \sigma_1
+\lambda_6 \sigma_2+ \frac{m_q d_1 d_2^2}{(d_1\sigma_1+d_2\sigma_2)^2}\Big)
V^{(2)\sigma}_{22}\!.\label{ineq3}
\end{eqnarray}
This last inequality is a necessary condition that has to be satisfied by the model at zero
chemical potential for it to be potentially capable of yielding SPB. Evidently, this inequality must
hold across the critical point in order that $\partial_\mu \rho^2 > 0, \partial_\mu (\pi^0)^2 > 0$.

\subsection{Second variations of effective potetial in the SPB phase: character of the phase transition in the chiral limit}
Once a condensate for $\pi^{0}_2$ appears spontaneously the vector $SU(2)$ symmetry is broken to
$U(1)$ and two charged $\Pi$ mesons are expected to possess zero masses as dictated by the Goldstone theorem.
For simplicity let us  consider zero temperature. In the chiral limit the matrix of second
variations in essential variables $\sigma_1, \sigma_2, \pi_2^0$ has the rank 3, $ \hat V^{(2)} = \left(V^{(2)}_{mn}\right); \, m,n = 1,2,0,$ where the index '0' is engaged for variation of the neutral pseudoscalar field $\pi^{0}_2$.  This matrix reads
\begin{eqnarray}
&&\frac12 V^{(2)\sigma}_{11} = -\Delta + 6\lambda_1 \sigma_1^2  +
3\lambda_5 \sigma_1 \sigma_2\nonumber\\ && + (\lambda_3 + \lambda_4)  \sigma_2^2 + (\lambda_3
- \lambda_4) \rho^2 \nonumber\\ &&
+  {\cal N}\left[\mu \sqrt{\mu^2-\sigma_1^2}-
3\sigma_1^2\ln{\frac{\mu+\sqrt{\mu^2-\sigma_1^2}}{\sigma_1}}\right] \equiv
\frac12 {\cal V}_{11},
\nonumber\\
&&V^{(2)\sigma}_{12} =  3 \lambda_5 \sigma_1^2  + 4  (\lambda_3
+ \lambda_4) \sigma_1 \sigma_2\nonumber\\
&& + 3\lambda_6  \sigma_2^2 + \lambda_6 \rho^2
\equiv {\cal V}_{12},
\nonumber\\
&&\frac12 V^{(2)\sigma}_{22} =-  \Delta  +  (\lambda_3 + \lambda_4)
\sigma_1^2 + 3 \lambda_6  \sigma_1 \sigma_2 \nonumber\\ &&+ 6 \lambda_2 \sigma_2^2 +
2\lambda_2 \rho^2  \equiv \frac12 {\cal V}_{22},\label{secvarpic1}\end{eqnarray}
\begin{eqnarray}&& V^{(2)\sigma\pi}_{10} = \Big( 4 (\lambda_3 - \lambda_4)\sigma_1 +
2\lambda_6\sigma_2 \Big) \rho \equiv {\cal V}_{10} \rho ,\nonumber\\
 &&V^{(2)\sigma\pi}_{20} = \Big( 2 \lambda_6 \sigma_1 + 8\lambda_2\sigma_2\Big)
\rho \equiv {\cal V}_{20} \rho  ,
\nonumber\\
&&\frac12 V^{(2)\pi}_{00} = 4 \lambda_2\rho^2 \equiv \frac12 {\cal V}_{00}
\rho^2 . \label{secvarpic}
\end{eqnarray}
We notice that the second variation of charged pseudoscalar fields $\pi_2^\pm$ vanishes $V^{(2)\pi}_{\pm\mp} = 0 $ and therefore these fields are massless Goldstone bosons.

Now we are able to check the character of phase transition. The qualitative behavior of the order parameters: dynamical mass $\sigma_j(\mu)$ and parity-odd condensate $\rho(\mu)$, is shown on Fig.2.  It is justified when using consistently equations
\gl{mg2}, \gl{var1} and the condition \gl{creq1m} in the SPB phase. Then one obtains the differential
equations on functions $\sigma_j(\mu),\ \rho(\mu)$, following the same strategy as for \gl{eqmu},
\begin{eqnarray}
&&\partial_\mu\sigma_1 = - 4 {\cal N} \sigma_1\sqrt{\mu^2-\sigma_1^2}
\frac{{\cal V}_{22}{\cal V}_{00}- {\cal V}^2_{20}}{{\rm Det}\hat{\cal V}} < 0 ,\nonumber\\
&&\partial_\mu\sigma_2 = - 4 {\cal N} \sigma_1\sqrt{\mu^2-\sigma_1^2}
\frac{{\cal V}_{10}{\cal V}_{20}- {\cal V}_{12}{\cal V}_{00}}{{\rm Det}\hat{\cal
V}},\nonumber\\
&& \partial_\mu\rho = - 4 {\cal N} \sigma_1\sqrt{\mu^2-\sigma_1^2}
\frac{{\cal V}_{12}{\cal V}_{20}- {\cal V}_{10}{\cal V}_{22}}{{\rm Det}\hat{\cal
V}}. \label{eqmurho}
\end{eqnarray}
The last derivative must be positive in order to generate parity breaking
and this is guaranteed by the inequality \gl{ineq3}.
\begin{figure}
[tbp]
\begin{minipage}{\columnwidth}
\hspace{-1.5cm}
\includegraphics[width=1.5\textwidth]{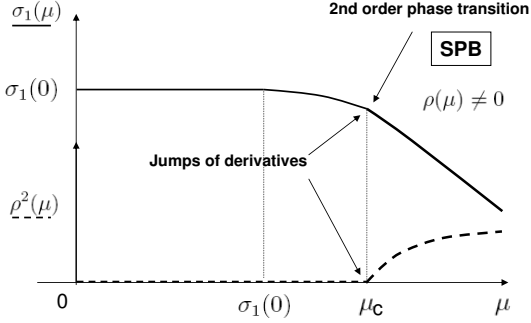}
\end{minipage}
\vspace{-3cm}
\caption{\sl The SPB phase transition of second order: the dashed line depicts the SPB breaking phase
and the solid line stands for the v.e.v. of "dynamical" mass. The plot is only qualitative.}
\end{figure}
Let us compare the derivatives of the dynamic mass $\sigma_1$ across the phase transition point.
For $\mu \rightarrow \mu_{crit} -i0$ its derivative is given by \gl{eqmu} and for $\mu \rightarrow \mu_{crit} +i0$
it is given by \gl{eqmurho}.

Their difference reads
\begin{eqnarray}
&&\partial_\mu\sigma_1\Big|_{\mu_{crit} +i0}-
\partial_\mu\sigma_1\Big|_{\mu_{crit} -i0}\nonumber\\
&& = - 4 {\cal N}
\sigma_1\sqrt{\mu^2-\sigma_1^2} \left\{
\frac{{\cal V}_{22}{\cal V}_{00}- {\cal V}^2_{20}}{{\rm Det}\hat{\cal V}} -
\frac{V^{(2)\sigma}_{22}}{{\rm Det}\hat V^{(2)\sigma}}\right\}\nonumber\\
&&= - 4 {\cal N} \sigma_1\sqrt{\mu^2-\sigma_1^2}\frac{\big({\cal V}_{10}
V^{(2)\sigma}_{22}- {\cal V}_{20} V^{(2)\sigma}_{12}\big)^2}{{\rm Det}\hat{\cal
V}\, {\rm Det}\hat V^{(2)\sigma}} < 0,
\end{eqnarray}
provided that both determinants are positive (they determine the spectrum of meson masses squared )
and inequality \gl{ineq3} holds across. Thus the dynamic mass derivative is discontinuous and the
phase transition is of second order.

\subsection{Inclusion of current quark masses}
For non-vanishing current quark masses the deviation linear in $m_q$  in the parity breaking phase affects
also the pseudoscalar parameters $\langle\rho\rangle = \rho(\mu)$ and $\langle\pi^0\rangle = \pi^0(\mu)$. After usage of equation \gl{vevpi} one finds
\begin{eqnarray}
&&\hat\Delta
\equiv\left(\begin{array}{c}\Delta\sigma_1(m_q)\\\Delta\sigma_2(m_q)\\
\Delta\rho (m_q)\end{array}\right) \simeq 2 m_q
\left(\begin{array}{c}\Delta_1\\\Delta_2\\ \frac{\Delta_0}{\rho}
\end{array}\right);\\
&&\tilde\Delta\equiv
\left(\begin{array}{c}\Delta_1\\\Delta_2\\ \Delta_0 \end{array}\right)\no&&=
\Big(\hat{\cal V}\Big)^{-1}\cdot\,
\left(\begin{array}{c}d_1\\d_2\\ \frac{d^2_2}{d_1 \sigma_1 + d_2 \sigma_2 }
\end{array}\right)\, \frac{d_1\sigma_1+d_2\sigma_2}{\sqrt{d_2^2 \rho^2 +
(d_1\sigma_1+d_2\sigma_2)^2}} ,\nonumber
\end{eqnarray}
where in order to keep the leading order all parameters must be taken in the chiral limit.
As to the v.e.v. of neutral pion field it does not need any mass corrections to the leading order and
must be taken from the mass independent eq. \gl{mg4}
\begin{equation}
\pi^0 =
-\mbox{arctan}\left(\frac{d_2\rho}{d_1\sigma_1+d_2\sigma_2}\right) .
\end{equation}
Accordingly, the mass corrections to the matrix of second variation $\Delta\hat V^{(2)}$, equation
\gl{secvarpic}, takes the form
\begin{eqnarray}
&&V^{(2)\sigma}_{jl} (m_q)
= V^{(2)\sigma}_{jl} (0) +\sum_{m = 1,2,0} \partial_m \left(
V^{(2)\sigma}_{jl} \right) \hat\Delta_m,\label{sigmajl} \\
&&V^{(2)}_{\sigma_j\pi_2^0} (m_q)  = V^{(2)\sigma\pi}_{j0}+\sum_{m = 1,2,0}
\partial_m \left( V^{(2)\sigma\pi}_{j0} \right) \hat\Delta_m, \label{sigmapi2}\\
&&\left(\partial_m\right) \equiv
\left(\partial_{\sigma_1}, \partial_{\sigma_2},
\partial_{\rho}\right);\no
&&V^{(2)}_{\sigma_1\pi_1^0} =
\frac{2m_q d_1}{F_0}\sin\frac{\pi^0}{F_0}, \quad
V^{(2)}_{\sigma_2\pi_1^0} =
\frac{2m_q d_2}{F_0}\sin\frac{\pi^0}{F_0}, \label{sigmapi1}\end{eqnarray}
\begin{eqnarray}
&&\frac12
V^{(2)\pi^0}_{11} =
m_q\Biggl(\frac{d_1\sigma_1+d_2\sigma_2}{F_0^2}\cos\frac{\pi^0}{F_0}
\no &&\phantom{\frac12
V^{(2)\pi^0}_{11} =}-\frac{d_2\rho}{F_0^2}\sin\frac{\pi^0}{F_0}\Biggr), \label{pi11}\\
&&V^{(2)\pi^0}_{12} =
\frac{2m_q d_2}{F_0}\cos\frac{\pi^0}{F_0}, \label{pi12}\\
&&\frac12
V^{(2)\pi^0}_{22} = 4 \lambda_2\rho^2 (0) + 16 m_q \lambda_2 \Delta_0\no &&\phantom{V^{(2)\pi^0}_{22} =}
+\frac{m_q d_2^2}{d_1\sigma_1+d_2\sigma_2}\cos\frac{\pi^0}{F_0},
\label{pi22}\\
&&V^{(2)}_{\pi^{+}_1\pi^{-}_1} =- 2
m_q\frac{d_2\rho}{(\pi^0)^2}\sin\frac{\pi^0}{F_0} \no &&\phantom{V^{(2)}_{\pi^{+}_1\pi^{-}_1}}= 2
m_q
\frac{d_1\sigma_1+d_2\sigma_2}{(\pi^0)^2}\frac{\sin^2\frac{
\pi^0}{F_0}}{\cos\frac{\pi^0}{F_0}},\label{pipm11}
\\
&&V^{(2)}_{\pi^{+}_1\pi^{-}_2} = V^{(2)}_{\pi^{+}_2\pi^{-}_1} =
\frac{2m_qd_2}{\pi^0}\sin\frac{\pi^0}{F_0}
,\label{pipm12} \\
&&V^{(2)}_{\pi^{+}_2\pi^{-}_2} =
\frac{2m_qd_2^2}{d_1\sigma_1+d_2\sigma_2}\cos\frac{\pi^0}{F_0},
\label{pipm22}
\end{eqnarray}
where the r.h.s. are evaluated with the help of eqs. \gl{mg1}-\gl{mg4}, \gl{creq1m} - \gl{creq2m} and
the v.e.v's for $\sigma_j, \rho$ are taken in the chiral limit. We notice that
convexity around this minimum implies that all diagonal elements are non-negative. This gives
positive masses for two scalar and four pseudoscalar mesons, whereas the doublet
of charged  of $\pi$ mesons remains massless. The latter can be easily checked from the vanishing
determinant of the last matrix $V^{(2)}_{\pi^{+}_j\pi^{-}_l}$ in eqs. \gl{pipm11}-\gl{pipm22}.
Of course, quantitatively the mass spectrum can be obtained only after kinetic terms are properly normalized.

If the soft breaking of chiral symmetry occurs only in the $H_1$ channel, $d_2=0$ then it follows from
eq. \gl{mg4} that light pions do not condense $\langle\pi^0\rangle = 0$ and do not mix with other states as
the off-diagonal matrix elements \gl{sigmapi1}, \gl{pi12} and \gl{pipm12} vanish. The second pair of charged
pseudoscalars $\pi^{\pm}_2$ becomes massless manifestly.

\section{Kinetic terms in two-multiplet $\sigma$ model}
In this Section we examine the fluctuations around the constant solutions of the mass-gap
equations \gl{var1},\gl{mg2}, \gl{mg4},\gl{mg3}, and introduce appropriate notations for
the fluctuations $\Sigma_j,\hat\Pi$ around  v.e.v.'s $\langle\sigma_j\rangle, \langle\rho\rangle$ so that
$\sigma_j \equiv \langle\sigma_j\rangle+ \Sigma_j$,$\hat\pi_1 = \tau_3 \langle\pi^0\rangle + \hat\pi$, $\hat\pi_2 = \tau_3 \langle\rho\rangle + \hat\Pi$.
These v.e.v's $\langle\sigma_j\rangle, \langle\rho\rangle$ must be used in all previous relations for the second variation of the
potential. In calculations of the kinetic term we retain only the terms in the chiral limit
keeping our interest to the masses of scalar and pseudoscalar mesons at the leading order of the
expansion in current quark masses, $m_q$. Thus in the kinetic terms we take $\langle \pi^0\rangle \simeq 0$
according to eqs. \gl{mg4},\gl{mg3}.

\subsection{General form of kinetic terms from chiral symmetry}
Once we have fixed the interaction to quark matter we are not free in the choice
of the kinetic term for scalar fields. Namely one cannot rotate two fields and rescale the field
$H_1$ without changes in the chemical potential driver \gl{potmu}. However the rescaling of the field
$H_2$ is possible at the expense of an appropriate redefinitions of other coupling constants and this
freedom can be used to fix one of the constants which appear in the kinetic term. Thus we take the
general kinetic term symmetric under $SU(2)_L\times SU(2)_R$ global rotations to be
\begin{equation}
{\cal
L}_{kin} = \frac14
\sum\limits^2_{j,k =1} A_{jk} \tr{\partial_\mu H^\dagger_j
\partial^\mu H_k} .
\end{equation}
With the chiral parameterization \gl{chipar} one can separate the bare Goldstone boson action,
\begin{eqnarray}
&&{\cal L}_{kin} = \frac12 \sum\limits^2_{j,k =1} A_{jk}
\partial_\mu \sigma_j \partial^\mu \sigma_k\nonumber\\ &&
+ \frac14 \sum\limits^2_{j,k =1} A_{jk} \sigma_j \sigma_k \mbox{\rm tr}\Bigl\{\partial_\mu
U^\dagger \partial^\mu
U\Bigr\}\\ && + \frac12 i \sum\limits^2_{j =1} A_{j2}\mbox{\rm tr}\Bigl\{\sigma_j\Big(\xi^\dagger
(\partial_\mu\xi)^2
\xi^\dagger -
\partial_\mu\xi (\xi^\dagger)^2 \partial^\mu\xi\Big)\hat\pi_2 \nonumber\\ &&
- \sigma_j \xi^\dagger\partial_\mu U \xi^\dagger\partial^\mu \hat\pi_2 +
\partial_\mu \sigma_j
\xi^\dagger\partial^\mu U \xi^\dagger \hat\pi_2\Bigr\}\nonumber\\ && + \frac14 A_{22} \mbox{\rm tr}\Bigl\{
\partial_\mu\hat\pi_2
\partial^\mu \hat\pi_2 - 2 \partial_\mu\xi \xi^\dagger \hat\pi_2 \xi^\dagger
\partial_\mu\xi
\hat\pi_2 \nonumber\\ && - (\partial_\mu\xi \xi^\dagger \partial^\mu\xi
\xi^\dagger +
\xi^\dagger\partial_\mu\xi
\xi^\dagger \partial^\mu\xi)(\hat\pi_2)^2 \nonumber\\ &&  + [\xi^\dagger,
\partial_\mu\xi][\hat\pi_2, \partial^\mu
\hat\pi_2]\Bigr\} . \nonumber
\end{eqnarray}
After selecting out the v.e.v. $\langle H_j\rangle = \langle\sigma_j\rangle$ let us explore the kinetic part quadratic in fields. We expand $U = 1 + i
\hat\pi/F_0 +\cdots,\ \xi = 1 + i \hat\pi/2F_0 +\cdots$ and use the notations defined at the beginning of this Section.
Then the quadratic part looks as follows
\begin{eqnarray}
{\cal L}^{(2)}_{kin} &=& \frac12 \sum\limits^2_{j,k =1} A_{jk}\Biggl[
\partial_\mu \Sigma_j \partial^\mu \Sigma_k \no&&+\frac{1}{F_0^2} \langle\sigma_j\rangle
\langle\sigma_k\rangle\partial_\mu \pi^a \partial^\mu \pi^a\Biggr] \label{kine2}
\\&&+\frac{1}{F_0}\sum\limits^2_{j =1} A_{j2}\Biggl[- \langle\rho\rangle
\partial_\mu \Sigma_j \partial^\mu \pi^0 +
\langle\sigma_j\rangle  \partial_\mu \pi^a \partial^\mu \Pi^a\Biggr] \no&&
+ \frac12 A_{22}\Biggl[ \frac{\langle\rho\rangle^2}{F^2_0}\partial_\mu \pi^0 \partial^\mu
\pi^0 + \partial_\mu
\Pi^a \partial^\mu \Pi^a\Biggr] , \nonumber
\end{eqnarray}
which shows manifestly the mixture between bare pseudoscalar
states and, in the SPB phase, also between scalar and pseudoscalar states.

Let us define
\begin{equation}
F_0^2 =  \sum\limits^2_{j,k =1} A_{jk} \langle\sigma_j\rangle
\langle\sigma_k\rangle ,\quad \zeta \equiv
\frac{1}{F_0} \sum\limits^2_{j =1} A_{j2} \langle\sigma_j\rangle . \label{norm1}
\end{equation}

\subsection{Parity-symmetric phase\label{pimass}}
In the symmetric phase $\langle\rho\rangle = 0, \, \hat\pi_2 = \hat\Pi$ one diagonalizes by shifting the pion field
\begin{eqnarray}
&&\pi^a = \tilde\pi^a -
\zeta \Pi^a , \label{shift}\\
&&{\cal L}^{(2)}_{kin,\pi} = \frac12 \partial_\mu \tilde\pi^a \partial^\mu
\tilde\pi^a + \frac12
(A_{22} - \zeta^2)\partial_\mu \Pi^a \partial^\mu \Pi^a , \no&& A_{22} -
\zeta^2 =
\frac{\langle\sigma_1\rangle^2 \mbox{\rm  det} A}{F_0^2} > 0 .
\end{eqnarray}
Taking into account the modification of the matrix of second variations \gl{pi11}-\gl{pipm22}
after shifting \gl{shift} one finds the masses of light and heavy pseudoscalars to the leading
order in current quark mass
\begin{eqnarray}
&&(\tilde V^{(2)\pi}_{11})^{ab} =
(V^{(2)\pi}_{11})^{ab}\nonumber\\
&&= \delta^{ab}2 m_q\frac{d_1\langle\sigma_1\rangle +d_2\langle\sigma_2\rangle }{F_0^2} =
\delta^{ab} m_\pi^2,
\nonumber\\
&&(\tilde V^{(2)\pi}_{12})^{ab} =(V^{(2)\pi}_{12})^{ab} - \zeta
(V^{(2)\pi}_{11})^{ab}\nonumber\\
&& = \delta^{ab} 2 m_q \left(\frac{d_2}{F_0} - \zeta
\frac{d_1\langle\sigma_1\rangle +d_2\langle\sigma_2\rangle}{F_0^2}\right),
\nonumber\\
&&(\tilde V^{(2)\pi}_{22})^{ab} =\left((V^{(2)\pi}_{22})^{ab} - 2\zeta
(V^{(2)\pi}_{12})^{ab} + \zeta^2 (V^{(2)\pi}_{11})^{ab}\right)\nonumber\\
&&=
\delta^{ab}2 \Big(- \Delta +(\lambda_3 - \lambda_4) \langle\sigma_1\rangle ^2 + \lambda_6
\langle\sigma_1\rangle \langle\sigma_2\rangle \nonumber\\
&& +
2\lambda_2 \langle\sigma_2\rangle ^2  - \zeta  m_q \frac{2 d_2}{F_0} + \zeta^2
m_q\frac{d_1\langle\sigma_1\rangle +d_2\langle\sigma_2\rangle}{F_0^2}\Big)
\nonumber\\&&= \delta^{ab} ((A_{22} - \zeta^2)) m_{\Pi}^2 + {\cal O}(m_q^2).
\label{secvar22}
\end{eqnarray}
at the leading order in $m_q$ because it is assumed that $m_{\Pi} \gg m_\pi$ far below the P-breaking transition point in chemical potential (see Fig. 3).
We notice that in this region the off-diagonal element does not make any influence.
\begin{figure}
[tbp]
\begin{minipage}{\columnwidth}
\hspace{-1.7cm}
\includegraphics[width=1.5\textwidth]{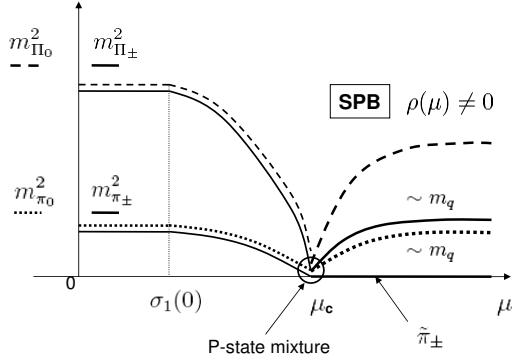}
\end{minipage}
\vspace{-3cm}
\caption{\sl Masses of pseudoscalar states in P-symmetric and in SPB phases. The light and
heavy charged pseudoscalars are depicted with solid lines, the dotted line corresponds to the neutral
light pseudoscalar and the dashed line stands for the neutral heavy one. The plot is only qualitative.}\nonumber
\end{figure}
At the point of the SPB phase transition one has to impose the condition \gl{creq1m} which leads to
\be
(\tilde V^{(2)\pi}_{22})^{ab}
=
\frac{\delta^{ab} 2m_q}{d_1\langle\sigma_1\rangle +d_2\langle\sigma_2\rangle} \Big(d_2 - \zeta \frac{d_1\langle\sigma_1\rangle +d_2\langle\sigma_2\rangle}{F_0} \Big)^2 .
\label{secvar222}
\ee
Evidently the determinant of $(\tilde V^{(2)\pi})$ vanishes and one reveals zero modes for all three
pion states, one neutral and two charged. They represent the true Goldstone modes (in the
limit of exact isospin symmetry $m_u=m_d$). At the P-breaking transition point $\langle\rho\rangle = 0$, when taking into
account the normalization  of kinetic terms \gl{kine2} with the definitions \gl{norm1} one finds the
values of three massive modes
\ba
&&m^2_{\Pi} = 2m_q \frac{A_{11} d^2_2 - 2 A_{12} d_1d_2
+ A_{22} d_1^2}{ \mbox{\rm det}A (d_1\langle\sigma_1\rangle +d_2\langle\sigma_2\rangle)}\nonumber\\
&& =
2 m_q \frac{(\sum_{j,k = 1,2} d_j [A^{-1}]_{jk} d_k)}{\sum_{j,k = 1,2} d_j  {\langle\sigma_j\rangle}} . \label{spbpi}
\ea

Thus in the chiral limit, at the phase transition point one reveals six zero modes and beyond
the chiral limit only three ones (see Fig.3).

\subsection{Masses of light states in SPB phase}
In the SPB phase the situation is more involved: pseudoscalar states mix with scalar ones. In
particular, the diagonalization of kinetic terms is different for neutral and charged pions because
the vector isospin symmetry is broken: $SU(2)_V \rightarrow U(1)$. Namely
\begin{eqnarray}  &&\tilde\pi^\pm =
\!\!\pi^\pm + \zeta \Pi^\pm ,
\no &&\!\!\tilde\pi^0 = \pi^0 + \frac{F_0^2}{F_0^2 + A_{22} \langle\rho\rangle^2} \Big(\zeta
\Pi^0 -
\frac{\langle\rho\rangle}{F_0}\sum\limits^2_{j =1} A_{j2}\Sigma_j\Big) .
\label{diagspb}
\end{eqnarray}
In this way SPB induces mixing of both massless and heavy neutral pions with scalars. The
(partially) diagonalized kinetic term has the following form
\begin{eqnarray}
&&{\cal L}^{(2)}_{kin} = \partial_\mu
\tilde\pi^\pm
\partial^\mu \tilde\pi^\mp + \frac12 \Big( 1 +\frac{A_{22} \langle\rho\rangle^2}{F^2_0}
\Big)\partial_\mu
\tilde\pi^0 \partial^\mu \tilde\pi^0 \no&& +  (A_{22} - \zeta^2)\partial_\mu \Pi^\pm
\partial^\mu \Pi^\mp \nonumber\\ &&
+\frac12 (A_{22}- \frac{F_0^2}{F_0^2 + A_{22} \langle\rho\rangle^2}\zeta^2)\partial_\mu
\Pi^0 \partial^\mu
\Pi^0 \nonumber\\ &&+ \frac12 \sum\limits^2_{j,k =1} \frac{A_{jk} F_0^2 + \langle\rho\rangle^2
\mbox{\rm  det} A
\delta_{1j}\delta_{1k}}{F_0^2 + A_{22} \langle\rho\rangle^2}\partial_\mu \Sigma_j \partial^\mu
\Sigma_k \nonumber\\ && -\frac{F_0
\langle\rho\rangle}{F_0^2 + A_{22} \langle\rho\rangle^2}\zeta\partial_\mu \Pi^0 \sum\limits^2_{j =1}
A_{j2}\partial^\mu \Sigma_j.
\end{eqnarray}
We see that even in the massless pion sector the isospin breaking $SU(2)_V \rightarrow U(1)$
occurs: neutral pions become less stable with a larger decay constant. Another observation is that in
the charged meson sector the relationship between massless $\pi$ and $\Pi$ remain the same as in the
symmetric phase.

Beyond the chiral limit  one can derive the masses of the lightest pseudo-goldstone states.
When $\langle\rho\rangle \gg m_q$ then in the mass matrix \gl{sigmajl}-
\gl{pipm22} the heavy mass parts \gl{sigmajl}, \gl{sigmapi2}, \gl{pi22} and the light
mass ones \gl{pi11}, \gl{pipm11} - \gl{pipm22} combine into an approximately block diagonal form
with additional off-diagonal elements \gl{sigmapi1} and \gl{pi12}, proportional to $m_q$.
The latter leads to factorization of the light pseudoscalar meson sector from the heavy meson one
to the order of $m_q^2$. Thus neglecting the mixture of heavy and light states one deals with the light
sector built of \gl{pi11}, \gl{pipm11} -\gl{pipm22} which after diagonalizing the kinetic term
by \gl{diagspb} projected on the light state sector gives the light pseudoscalar masses
\begin{eqnarray}
&&m^2_{\tilde\pi_0} = 2
m_q \Big( 1 +\frac{A_{22} \langle\rho\rangle^2}{F^2_0}\Big)^{-1} \no&&\times \left(\frac{d_1\langle\sigma_1\rangle +d_2\langle\sigma_2\rangle}{F_0^2}\cos\frac{\langle\pi^0\rangle}{F_0}
-\frac{d_2\langle\rho\rangle}{F_0^2}\sin\frac{\langle\pi^0\rangle}{F_0}\right),\nonumber\\
&&m^2_{\tilde\pi^\pm } = 0,\label{masspseud}\\
&&m^2_{\Pi^\pm } =  2 m_q \dfrac{\cos\frac{\langle\pi^0\rangle}{F_0
}}{(A_{22} - \zeta^2)(d_1\langle\sigma_1\rangle +d_2\langle\sigma_2\rangle)}\no&&\times \left(d_2- \zeta
\frac{d_1\langle\sigma_1\rangle +d_2\langle\sigma_2\rangle}{\langle\pi^0\rangle}\tan\frac{\langle\pi^0\rangle}{F_0} \right)^2 .\
\end{eqnarray}

Thus in the SPB one finds two massless charged pseudoscalars and three light pseudoscalars with masses
linear in the current quark mass (see Fig.3). These equations represent the generalization of the Gell-Mann-Oakes-Renner
relation in the phase with broken parity.

We notice that the masses of neutral and charged pseudoscalars do not coincide in the
well developed SPB phase, just realizing the spontaneous breaking of isospin symmetry.
One can also guess that the manifest breaking of $SU(2)$ symmetry due to
different masses of $u$ and $d$ quarks will supply the Goldstone bosons
$\tilde\pi^\pm$ with tiny masses proportional to the difference $m_u - m_d$.

\section{Description of baryon matter in the mean-field approach}
When keeping in mind QCD we assume that the quark matter is equivalent to nuclear matter when their average baryon densities coincide,
at least in what respects meson properties. One could also think about techniquarks and the two multiplets of composite Higgs mesons.

Thermodynamical characteristics of such a matter are the pressure, $P$, and the energy density, $\varepsilon$.
The pressure is determined in the presence of chemical potential by \gl{press}, defined for $\sigma_j$
satisfying the mass gap equation. The pressure at zero nuclear density must vanish.
In this case the energy and baryon densities are related to the pressure as follows
\begin{equation}
\varepsilon = - P +N_c \mu \varrho_B;\ \partial_\mu P = N_c \varrho_B;\
\partial_\varrho \varepsilon = N_c \mu . \label{enden1}
\end{equation}
The direct connection between energy density and pressure reads
\begin{equation}
P = \varrho_B^2 \partial_{\varrho}\left(\frac{\varepsilon}{\varrho_B}\right) .
\end{equation}
Evidently the energy per baryon has an extremum when the pressure vanishes. Since the pressure is an
increasing function of the density as we have seen, obviously vanishing at zero density, and
infinite nuclear matter is stable (thus implying zero pressure) the phase diagram in the $P,\varrho_B$ plane
is necessarily discontinuous with values of density in the interval
$(0,\varrho_0)$ not corresponding to equilibrium states ($\varrho_0$ is nuclear matter density). We
will see below how this is realized in our model.

\subsection{On the way to stable nuclear matter}
Our model consisting of two scalar isomultiplets is still somewhat too simple in one aspect. The
stabilization of nuclear matter requires not only attractive scalar forces (scalars) but
also repulsive ones (vector-mediated). Conventionally \cite{walec}, the latter ones are associated to
the interactions mediated by the iso-singlet vector $\omega$ meson. Let us supplement our action with
the free $\omega$ meson lagrangian and its coupling to quarks
\begin{equation}
\Delta {\cal L}_\omega = - \frac14 \omega_{\mu\nu}\omega^{\mu\nu} +\frac12 m_\omega^2
\omega_\mu \omega^\mu - g_{\omega\bar q q} \bar q \gamma_\mu \omega^\mu q, \label{omega}
\end{equation}
with a coupling constant $g_{\omega\bar q q}\sim {\cal O} (1/\sqrt{N_c})$. After bo\-so\-ni\-za\-tion of QCD or QCD-like theories,
on symmetry grounds, any vector field interacts with scalars in the form of commutator and therefore
$\omega_\mu $  does not show up in the effective potential $H_j$ fields to the lowest order. However
in the quark lagrangian the time component $\omega_0$ interplays with the chemical potential and it is
of importance to describe the dense nuclear matter properties. Let us assign a constant v.e.v. for
this component $g_{\omega\bar q q} \langle\omega_0 \rangle \equiv \tilde\omega$. Then one needs to
compute the modification of the effective potential due to the replacement
$\mu \rightarrow \mu + \tilde\omega\equiv \tilde\mu$. The variable $\tilde\omega $, and accordingly
$\tilde\mu$, is dynamical and in addition appears quadratically in the mass term in \gl{omega} which
reads
\ba
&&\Delta V_\omega = - \frac12 m_\omega^2\langle \omega_0^2\rangle = -\frac12
\frac{(\tilde\mu - \mu)^2}{G_\omega},\no&& G_\omega \equiv
\frac{g_{\omega\bar q q}^2}{m_\omega^2}\simeq {\cal O} (\frac{1}{N_c}) .\label{effomega}
\ea
The term \gl{effomega} supplements the effective potential \gl{effmu}: $\widetilde V_{\mbox{\rm eff}, \omega}(\mu) = \widetilde V_{\mbox{\rm eff  }}(\tilde \mu) + \Delta V_\omega (\mu, \tilde\mu)$. Correspondingly the
extremum condition for the variation of the variable $\tilde\mu $ involves both the scalar part of the
effective potential \gl{effmu} and the vector one \gl{effomega} and due to \gl{varrho} takes the
following form
\begin{equation}
N_c \varrho_B (\mu) = \frac{N_c N_f}{3\pi^2} p_F^3 (\tilde\mu) =
\frac{\mu - \tilde\mu}{G_\omega}, \label{physmu}
\end{equation}
from this one finds $\tilde\mu(\mu)$ after solving the mass-gap equations \gl{var1},
\gl{mg2} and \gl{mg3}.

Finally the extended effective potential at a minimum reads
\ba
&&\widetilde V_{\mbox{\rm eff}, \omega
}(\mu) =- \frac12 \Delta \Big(\sigma_1^2(\tilde\mu) + \sigma_2^2(\tilde\mu) +
\langle\rho\rangle^2(\tilde\mu)\Big)\no&&  - \Big[\frac{{\cal N}}{3} \tilde\mu \Big(\tilde\mu^2 - \sigma_1(\tilde\mu)\Big)^{3/2}
\no&&+   G_\omega \frac{8{\cal N}^2}{9} \Big(\tilde\mu^2 - \sigma_1(\tilde\mu)\Big)^{3}\Big] \theta\Big(\tilde\mu -
\sigma_1(\tilde\mu)\Big) , \label{effmu+}
\ea
where $\langle\sigma_j\rangle = \sigma_j (\tilde\mu)$.
Let us define the v.e.v.'s of scalar field $\sigma_1$ in vacuum at the two minima as $\sigma_1^\ast(0) < \sigma^\sharp_{1}(0)$ . Let us select out the parameter subspace such that the minimum corresponding to $\sigma^\sharp_{1}(0), \sigma^\sharp_{2}(0)$ is lower than the another minimum at $\sigma^\ast_{1}(0), \sigma^\ast_{2}(0)$ .
Then for parity-even
matter $\langle\rho\rangle = 0$,  one seeks for the nuclear matter stability at a value of chemical potential $\tilde\mu_s$ with $\sigma_1^\ast (0) \leq \tilde\mu_s < \sigma^\sharp_{1}(0)$ . The corresponding baryon matter stability condition $\Delta P = P(\sigma_1^\sharp (0)) - P(\sigma_1^\ast (\tilde\mu_s)) = 0$, eq. \gl{press}, can be formulated as
\begin{eqnarray}
&&\Delta \Big((\sigma_1^\sharp (0))^2 + (\sigma_2^\sharp (0))^2-
(\sigma_1^\ast(\tilde\mu_s))^2 - (\sigma_2^\ast(\tilde\mu_s))^2\Big)\no && =\frac{N_cN_f}{6\pi^2} \tilde\mu_s p_F^3
(\tilde\mu_s) + G_\omega \frac{N_c^2 N_f^2}{9\pi^4} p_F^6 (\tilde\mu_s)\no &&= \frac{N_c}{2}
\tilde\mu_s \varrho_B (\mu_s) + G_\omega N_c^2 \varrho_B^2 (\mu_s), \label{stablenm}
\end{eqnarray}
taking into account \gl{effmu+}
and \gl{effomega} . Herein
$\tilde\mu_s$ is related to the physical value of $\mu_s$ by \gl{physmu}
and it is assumed that parity is not violated $\langle\rho\rangle = 0$. This relation represents the condition for
the formation of stable symmetric nuclear matter in result of first-order phase transition
\cite{walec}. It can be fulfilled by an appropriate choice of the vector coupling constant $G_\omega$
as typically the first term in the r.h.s. of \gl{stablenm} is smaller than the one on the l.h.s. The first order phase transition at the saturation point is illustrated on Fig.4.

At finite temperatures one has to modify the thermodynamic relations.
The modification of the effective potential due to $\omega$ mesons is given by \gl{effomega}. Thus
$\widetilde V_{\mbox{\rm eff}, \omega }(\mu,\beta) \equiv
\widetilde V_{\mbox{\rm eff}}(\tilde \mu,\beta) + \Delta V_\omega (\tilde\mu, \mu)$
which  should henceforth be used in all the previous thermodynamical formulae. The replacement
$\mu\to \tilde\mu$ makes all expectation values depend rather on $\tilde\mu$ which is
determined via the variation of $\widetilde  V_{\mbox{\rm eff }}$
\begin{equation}
\frac{\tilde\mu - \mu}{G_\omega} = - N_c \varrho_B \Big(\beta,\mu,\sigma_1\Big) =
\partial_{\tilde\mu}  \widetilde V_{\mbox{\rm eff  }}(\tilde\mu,\beta). \label{physmu1}
\end{equation}
\begin{figure}
[tbp]
\begin{minipage}{\columnwidth}
\hspace{-1.4cm}
\includegraphics[width=1.5\textwidth]{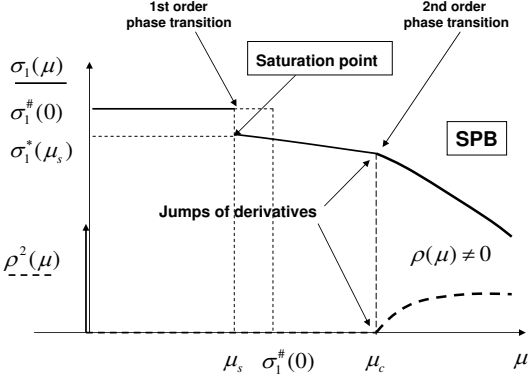}
\end{minipage}
\vspace{-2cm}
\caption{\sl Saturation point meets spontaneous parity breaking: at $\mu = \mu_s$ the pressures for the two solutions $\sigma^\sharp, \sigma^\ast$ become equal and the solutions interchange realizing the 1st order phase transition. At a larger chemical potential $\mu_c$ the 2nd order SPB phase transition occurs.  }\nonumber
\end{figure}
The saturation point at $\mu = \mu_s$ where nuclear matter forms is characterized by
the energy crossing condition for $P, T \not= 0$,
\begin{eqnarray}
&&\Delta \Big(\big(\sigma_1^\sharp\big)^2 + \big(\sigma_2^\sharp\big)^2-
\big(\sigma_1^\ast\big)^2 - \big(\sigma_2^\ast\big)^2\Big)\no && =
\frac{N_c}{2} \tilde\mu_s \Big(\varrho_B (\beta,\mu_s, \sigma_1^\ast)- \varrho_B
(\beta,\mu_s, \sigma_1^\sharp)\Big) \nonumber\\&&  + \frac12 T \Big(S(\beta,\mu_s, \sigma_1^\ast)-
S(\beta,\mu_s, \sigma_1^\sharp)\Big) \no &&+ G_\omega N_c^2 \Big(\varrho_B^2 (\beta,\mu_s,
\sigma_1^\ast)- \varrho_B^2 (\beta,\mu_s, \sigma_1^\sharp)\Big), \label{stablenm1}
\end{eqnarray}
where $\tilde\mu_s$ is related to the physical value of $\mu_s$ by equation \gl{physmu1} and
$\sigma_j^\sharp\equiv \sigma_j^\sharp(\tilde\mu_s, \beta);\ \sigma_j^\ast \equiv \sigma_j^\ast(\tilde\mu_s,\beta)$.

The latter relation represents the condition for the existence of symmetric nuclear matter.
It can  be always fulfilled by an appropriate choice of $G_\omega$.


\subsection{Saturation point meets spontaneous parity breaking}
Let us search for the domain of parameters in the  model providing the realization of both stable nuclear
matter and the regime of SPB.  The former is associated with a first-order phase transition and implies
the existence of two minima at zero chemical potential which are possibly moving when the chemical potential
increases. The highest, metastable minimum must start moving at chemical potentials $\mu$ smaller than the
value of the dynamical mass of the lowest minimum, $\sigma_1^\sharp$ and larger than the v.e.v. $\sigma_1^\ast$
at the highest minimum, \mbox{$\sigma_1^\ast\leq\mu < \sigma_1^\sharp$}. This metastable minimum may reach the lowest one
if the density and omega meson effects are taken into account. Then a first-order phase transition to
normal nuclear matter occurs when pressures become equal, eq. \gl{stablenm}.


In order to simplify our search we make a particular choice of $\lambda_{5}=0$ (not a reduction by \gl{lintrans}!).
In this case one of the solutions is $\sigma^{(0)}_2 = 0$ and $2\lambda_1 (\sigma^{(0)}_1)^2 =\Delta$ and it is a {\it minimum} as it follows from \gl{secvar}, \gl{secvar1} (in the chiral limit) provided that $\lambda_1 >0,
 (\lambda_3+\lambda_4)   > 2\lambda_1 $ .  When $\sigma^{(0)}_2 = 0$
a higher symmetry $Z_2 \times Z_2$ arises for the effective potential in the vicinity of such a
minimum as the contribution of the vertex with $\lambda_6$ into the second variation vanishes with
$\sigma^{(0)}_2 $. For $\sigma_2 \not= 0$ one can obtain eq. \gl{cubic} for the ratio
$t =\sigma_2/\sigma_1$. As it is analyzed in subsection 4.2,  it has, in general, one or three real
roots. For our purposes eq. \gl{cubic} must have three real solutions: one corresponding to a minimum $t^{(3)} >0$ and two
corresponding to saddle points $t^{(1)} < 0, t^{(2)} > 0$.   The inequality controlling the existence of three real
solutions is derived in subsection 4.2 from the analysis of the minimum of the polynomial \gl{cubic}.
Finally for a given solution $t^{(3)}$ one finds a unique solution for $\sigma_1 > 0$ from \gl{mg1} (or \gl{var1})
and \gl{mg2} .

Let us assume the minimum  with $\sigma_2^{(0)} = 0$ to be the higher one at zero chemical potentials $\sigma^{(0)}_1 \equiv \sigma^\ast_1$. For this
choice to be realized it is {\it sufficient} to fulfill the inequality $\sigma_1^{(0)} \equiv \sigma^\ast_1 < \sigma_1^{(3)}\equiv \sigma_1^\sharp$ .  It
turns out that in order to provide it one has to satisfy  the inequality
\begin{equation}
 2\lambda_2 t^2 +\frac32 \lambda_6 t +(\lambda_3 + \lambda_4 - 2\lambda_1) < 0, \label{ineqq1}
\end{equation}
which implies
 \be
 9\lambda^2_6 \geq 32\lambda_2(\lambda_3 +\lambda_4 - 2 \lambda_1);\quad 0< t^{(3)} < -\frac{3\lambda_6}{4\lambda_2} .
 \ee
 When at the critical value of $\mu = \mu_s < \sigma_1^\sharp$, the solution with $\sigma^\ast_2 = 0$ describes
a saturation point then the further evolution of the meson background for higher chemical potentials is
characterized by the following equation $\sigma_1^\ast (\mu^\ast)$
 \be
 2\lambda_1 (\sigma_1^\ast)^2 (\mu) = \Delta - {\cal N} {\cal A} (\sigma_1^\ast, \mu).
 \ee
As the last term is monotonously increasing with chemical potential the v.e.v of scalar field is decreasing.
Now we approach to the P-breaking regime and employ eq. \gl{creq1} at the expected phase transition point.
Its solution is
\be
\sigma^2_{1,c} = \frac{\Delta}{\lambda_3-\lambda_4} < (\sigma_1^\#)^2 = \frac{\Delta}{2\lambda_1} .
\ee
Thus the feasibility of spontaneous P-breaking depends on the realization of the inequality
$\sigma_{1,c} <\sigma^\ast_{1} <\sigma^\#_1$. The combination of the regime of nuclear matter saturation for normal baryon density and  of the P-breaking phase at a higher baryon density is qualitatively depicted on Fig.4.

Let us collect the inequalities providing the required convexity of the two minima
 and the very existence of both the stable nuclear matter and a
parity breaking phase for higher densities (see \cite{anes,aaes})
\begin{eqnarray}
&&
-\frac{3\lambda_6}{4\lambda_2}> t^{(3)} >\max\Big[-\frac{3\lambda_6}{8\lambda_2}; -\frac{4\lambda_4}{\lambda_6}\Big],\no&& \lambda_{1,2,3,4} > 0 ,\ \lambda_3 > \lambda_4,\
\frac32 \lambda_6^2 > 8 \lambda_2 \lambda_4> \lambda_6^2,\nonumber\\ && \Delta>0,\  (\lambda_3\pm\lambda_4)   > 2\lambda_1,\  (\lambda_3+\lambda_4)   > 2\lambda_2 ,
\end{eqnarray}
in addition to those ones derived above. For a more definite numerical estimation of these six constants associated to QCD
there is not at present  enough  experimental or phenomenological information.

\section{Confronting the two-multiplet model with  meson and nuclear matter phenomenology}
We assume that the quark matter is equivalent to nuclear matter when their average baryon densities coincide,
at least in what respects meson properties. Thus the two-multiplet model investigated in our paper could be exploited to explore baryon matter properties in the mean-field approach. The baryon matter normalization we will apply at the normal baryon density. The normal density of infinite nuclear matter \cite{walec} is $\varrho_0 \simeq 0.15\div
0.16$ fm$^{-3}$ that corresponds to the average distance $1.8 \div 1.9$ fm between nucleons in nuclear matter.
The two-multiplet model investigated in our paper contains a number of empirical parameters which at present are difficult to calculate directly from QCD. Instead following the assumption that QCD governs exactly the phenomenology of hadron physics one can attempt to derive these parameters and coupling constants from the very meson and nuclear matter experimental data.
In total we have three dim-2 vertices with mass-like parameters $\Delta_{ij}$, three normalization parameters for kinetic terms $A_{ij}$ and six coupling constants for dim-4 meson self-interaction $\lambda_k$. Beyond the chiral limit one has also the two vertices linear in current quark masses parameterized by $d_j$. At last in order to provide the first order phase transition to stable baryon matter one has to include the repulsive forces generated by $\omega$ meson with the relevant coupling constant $G_\omega$ in \eqref{effomega}. All together one has 15 constants to be found from spectral characteristics of mesons and stable baryon matter.

The reparameterization \eqref{lintr} of the scalar field $H_2$ discussed in Sec.4 allows to reduce the mass-like parameters to only one, $\Delta_{ij} \rightarrow \Delta \delta_{ij}$.
Thus 13 independent parameters must be fixed from hadron phenomenology.

The first source for determination of coupling constants and mass scales of the model comes from the mass spectrum of two lightest multiplets of scalar isoscalar and pseudoscalar isotriplet mesons. The pseudoscalar meson masses are known with a reasonable precision according to \cite{pdg}. In particular the heavy pion $\Pi$ mass starts from $\sim 1300$ MeV in the vacuum, at zero temperature and chemical potential. The situation with scalar meson masses is less definite. In the mass range below 2 GeV there might be at least four  mixed scalar mesons with a glueball among them. The decay constants are well measured for light pions but not so precisely known for heavy pions and scalars. But in principle one could have 4+4 = 8 experimental inputs from meson masses and decay constants in dominating decay channels.

The fitting of nuclear matter properties could give more inputs for determination of model parameters. Namely, the normal density of infinite nuclear matter \cite{walec} is $\varrho_0 \simeq 0.15\div
0.16$ fm$^{-3}$ that corresponds to the average distance $1.8 \div 1.9$ fm between nucleons in nuclear matter. At the saturation point $P=0$ and
\ba
&&\mu_N  = \frac{\varepsilon}{ \varrho_B} = \mbox{energy per baryon} \no&&= m_N - E_{bound}= (939 -
16) MeV = 923 MeV .
\ea
The quark matter chemical potential is defined as $ \partial_\varrho \varepsilon = \mu_N = N_c \mu $.
Therefore at the saturation point $\mu_s = 308 MeV$. Then from \gl{stablenm1} it can be established that if normal
nuclear matter is formed at the chemical potential $\mu_s \simeq 308 MeV$ then it can
stabilized by $\omega$ meson condensate with $G_\omega \sim (10 \div 15) GeV^{-2}$ in a qualitative agreement to
what is known from other model estimations \cite{vectorm}.

Evidently for a more definite numerical estimation of the entire set of 13 constants
there is not at present  enough  experimental or phenomenological information, although it can be shown that
the tentative values assumed  in \cite{anes} for $\lambda_1 \sim 0.15, \lambda_3 \sim 4, \Delta \sim 0.03 GeV^{-2}$
may lead to the occurrence of SPB at about three times normal nuclear densities.

Still we pay hopes to collect the required number of inputs from hadron phenomenology to falsify the realization of spontaneous parity breaking in dense baryon matter or vice versa the discovery of SPB in heavy ion collisions \cite{aaep} might give the missing data to fix the model parameters with a reasonable precision.

\section{Conclusions}
In this paper we followed the preliminary investigations in \cite{anes,aaes} and explored the issue of parity
breaking in dense baryon matter employing effective lagrangian techniques.
\begin{itemize}
\item Our effective lagrangian is a realization of the generalized
linear $\sigma$ model, but including the two lowest lying resonances in each
channel, those that are expected to play a role in this issue. It can be associated with QCD or QCD-like technicolor models and includes the vertices of soft breaking of chiral symmetry presumably generated by current quark masses. In this minimal model
condensation of one of the pseudoscalar fields can arise on the background of two-component scalar
condensate so that the chiral constant background cannot be rotated away by transformation of
two complex scalar multiplets preserving space parity. The use of
effective lagrangians is crucial to understand how would
parity breaking originating from a finite baryon density eventually reflect in hadronic physics.
\item We conclude that parity breaking is  a realistic possibility in
nuclear matter at moderate densities and non-zero current quark masses. To prove it we included a chemical potential for the quarks that corresponds to a finite density of
baryons and investigate the pattern of symmetry breaking in its presence. The necessary
and sufficient conditions (beyond the chiral limit) were found for a phase where parity is  spontaneously broken to exist. It also
extends to finite temperature although for large temperatures the hot pion gas corrections must be taken into account.
\item As a consequence of SPB a strong mixing between scalar and pseudoscalar states appears that translate
spontaneous parity breaking into meson decays. The mass eigenstates will decay both in odd and even
number of "pions" simultaneously.
\item At the very point of the phase transition leading to parity breaking one has
{\it six} massless pion-like states in the chiral limit and the {\it three} massless
states when the quark masses are taken into account. After crossing the phase transition, in the parity broken
phase, the massless charged pseudoscalar states remain as Goldstone bosons enhancing  charged pion
production, whereas the additional neutral pseudoscalar state becomes massive.
\item As a bonus we have gotten a rather good description
of several aspects of nuclear physics; in particular a good description of the physics associated to
the condensation transition where nuclear matter becomes the preferred solution. The model is rich enough to
provide the relevant characteristics while avoiding some undesirable properties of simpler models, such as
the chiral collapse (see Appendix A). Other nuclear properties such as (in)\-com\-pres\-si\-bilities are well described too (see Appen\-dix B).
\end{itemize}
We have presented our results trying to avoid as much as possible specific numerical values for the different
quantities and parameters. Not only is this procedure more general but also the logical connections are better
outlined. The main conclusion of our studies is that spontaneous parity breaking seems to  be a
rather generic phenomenon at finite density. It would be interesting to investigate how
this new phenomenon could modify the equation of state of neutron stars
(the density of such objects seems to be
about right for it). It is also mandatory to investigate in detail the appearance of a
SPB phase in heavy-ion collisions.

\section{Acknowledgments} This work was supported by Research
grant FPA2010-20807. AA and VA are partially supported by Grant RFBR 13-02-00127 and by
Saint-Petersburg State University research grant 11.38.660.2013.

\appendix
\section{Stable baryon matter without chiral collapse}
A viable model of dense baryon (quark/nucleon) matter must reveal the phase transition to a stable
bound state at the normal nuclear density $\varrho_B = \varrho_0 $ for infinite homogeneous
symmetric nuclear matter at the so called ``saturation point''. This phase
transition is believed to be of first order similar to the vapor condensation into liquid: from droplets
of heavy nuclei to a homogeneous nuclear liquid. However in simple quark models of the NJL
type \cite{buballa1} this phase transition (for vanishing current quark masses) goes  to
the chirally symmetric phase  with zero dynamical mass (zero v.e.v. of scalar fields), so called
``chiral collapse''. When it happens the typical baryon density is substantially larger that the
normal one $\varrho_{B,c} = 2.8 \varrho_0$. For this reason these simple models cannot be a reasonable guide to
phase transitions in very dense nuclear matter.

Let us examine under which  conditions  the saturation point in our model happens to
be at normal nuclear density and is not accompanied by the chiral collapse keeping the dynamical mass
different from zero. To analyze this problem we have to examine the pressure in cold ($T=0$), dense
baryon matter.

We remind that in so far as our system undergoes spontaneous CSB the effective potential \gl{veff}
does not reveal any minimum at the origin in variables  $\sigma_j$ (for $\mu=0$) and may have either
a saddle point,$$\det\left[\hat V^{(2)}\right](\sigma_j=0) < 0$$ or a maximum, $$\det\left[\hat
V^{(2)}\right](0) > 0,\ \tr{\hat V^{(2)}} (0) < 0 . $$ One has a positive definite matrix of second variations
\gl{secvar}, \gl{secvar1} of the effective potential in the vicinity
of a CSB solution, $$\det\left[\hat V^{(2)}\right](\sigma_j^\sharp) > 0;\ \tr{\hat
V^{(2)}} (\sigma_j^\sharp) > 0 .$$ It means that $$V_{\mbox{eff}}(\sigma_j^\sharp) < V_{\mbox{eff}}(0).$$

These properties allow us to guess that at some value of chemical potential $\mu_s < \sigma_1^\sharp$
and smaller values of v.e.v.'s for scalar fields $ \sigma_1^\ast < \mu_s$ the deficit in scalar background
energies on the left-hand side of \gl{stablenm} may be exactly compensated by contributions from the
nuclear density and omega-meson repulsion on the right-hand side so that $P(\sigma_{j}^\ast,\mu_s) = 0$
and the system undergoes a first-order phase transition to the stable quark (nuclear) matter.

Let us now prove that for a large variety of coupling constants admitting CSB (and SPB, see below) one
of the v.e.v. $\sigma_j^\ast \not=0$ in the chiral limit, and the chiral collapse is impossible.
Indeed, suppose that $\sigma_j^\ast =0$ at  $\mu^\ast < \sigma_1^\sharp$ where the pressure vanishes then
\begin{equation}
(\mu^\ast)^4 + \frac{8G_\omega {\cal N}}{3}(\mu^\ast)^6 =
- \frac{3}{{\cal N}}\left|V_{\mbox{\rm eff  }}(\sigma_j^\sharp,\mu = 0)\right|.
\end{equation}
In this case the second variation matrix for the effective potential \gl{secvar}, \gl{svar1} reads
\begin{eqnarray}
&&\frac12 V^{(2)\sigma}_{11} = -\Delta + {\cal N} (\mu^\ast)^2 , \no&&
 V^{(2)\sigma}_{12} = 0,\quad
 \frac12
V^{(2)\sigma}_{22} = -  \Delta . \label{secvar0}
\end{eqnarray}
In order to induce SPB one takes $\Delta > 0$ (see below). Then from \gl{secvar0} one finds that  for
any value of $\mu^\ast$ the second variation matrix is never positive definite
and one reveals either a saddle point or a maximum at a presumed saturation point whereas a maximum
remains for vacuum values with $\mu^\ast < \sigma_1^\sharp$.  As we have to guarantee the existence
of stable nuclear matter with normal baryon density we consider
 further on $\Delta > 0$.

It is instructive to reduce the two-multiplet sigma model to a one-multiplet lagrangian associated to
a NJL quark model. In relations \gl{secvar0} it simply corresponds to taking only
one matrix element $V^{(2)\sigma}_{11}$ to describe the behavior around extremum at the origin. Evidently,
there is always a value of $\mu^\ast$ for which it becomes positive and chiral collapse is
inevitable. But when comparing with our model one concludes that the reason for
appearance of chiral collapse is not the absence of confinement \cite{buballa1}
but the inefficiency of a one-channel linear sigma model in representing the
complicated chiral dynamics in hadronic physics.
\section{Proposal for (in)compressibilities: matching quark and nuclear matters }
The (in)compressibility in the quark matter must be  defined as $ K(\mu) =
 \partial_{\varrho_B} P $, where the derivative is made with the help of the function
$\varrho_B(\mu)$ given in (\ref{rhomu}). For a zero pressure state (such as stable nuclear matter)
this equals
\begin{equation}
K= \varrho_B^2
\partial^2_{\varrho_B}\left(\frac{\varepsilon}{\varrho_B}\right)\bigl|_{P=0},
\end{equation}
which must be positive since it corresponds to a minimum of the energy per baryon. In our model this is
indeed the case
\begin{eqnarray}
&&K(\mu)=  N_c\varrho_B \partial_{\varrho_B} \mu=\frac{ p_F^2
(\tilde\mu)}{\Big(\tilde\mu-\sigma_1\partial_{\tilde\mu} \sigma_1 (\tilde\mu)\Big)} +  9 G_\omega
\varrho_B (\mu) , \no&&\partial_{\tilde\mu}\sigma_1 = - 4 {\cal N}
\sigma_1\sqrt{\tilde\mu^2-\sigma_1^2} \, \frac{V_{\sigma_2\sigma_2}^{(2)}}{\det \hat V^{(2)}} < 0 ,
\label{compress}
\end{eqnarray}
where $N_c = 3$ is assumed and $\sigma_j = \sigma_j(\tilde\mu)$.

In order to adjust the properties of stable baryon matter in a quark description
we parameterize the (in)\-com\-pres\-sibi\-lity as follows:\,
$ K =  a N_c \varrho_B \partial_\varrho \mu\Bigl|_{P=0}$.
Let us show that the matching of quark and nuclear matter at the saturation point is provided by the
normalization factor $a = 9$ in meson-nucleon models and $a = 1$ for quark-meson models.
The incompressibility  must be positive (giving a minimum of the energy per
baryon) when a stable nuclear matter is formed at zero pressure.  The derivative of the dynamical mass
is given in \gl{compress} wherefrom it follows that the derivative of the pressure is always
positive. Thus if there is a solution with $\sigma_1^\ast \equiv \sigma_{1}(\mu^\ast)< \mu^\ast < \sigma_{1}^\sharp$
providing $P=0$, the phase transition emerges to the stable nuclear matter state.

In the terms of nuclear d.o.f.  one defines
\be
\varepsilon = - P + \mu_N \varrho_B .
\ee
At the saturation point $P=0$ and
\be
\mu_N  = \frac{\varepsilon}{ \varrho_B} = \mbox{energy per baryon} = m_N - E_{bound}.
\ee
The quark matter chemical potential is defined as $ \partial_\varrho \varepsilon = \mu_N = N_c \mu $.
 Let us use this point for the quark-hadron matching
\ba
&&\varrho_B(\mu_N) =  \varrho_B (\mu_s),\no&& p_{F,N}^2 = (\mu_N)^2 - (m_N)^2
\simeq p_{F,q}^2 =(\mu_s)^2 - (\sigma_1^\ast)^2 ,
\ea
if we neglect the vector meson shift $\tilde\mu \simeq\mu$. However the matching of densities does not
provide the equivalence of derivatives w.r.t. chemical potentials. Namely around the saturation point
\be
\frac{\partial p^2_F}{\partial \mu_N} \simeq 2 \mu_N = 2N_c \mu_s \not= \frac{\partial
p^2_F}{\partial \mu} \simeq 2\mu_s ,
\ee
where  (subdominant) derivatives of dynamical masses are neglected for a moment. Now let us try to
extend the matching to the (in)compressibilities, providing the correct $N_c$ factors. For hadron matter
\be
K_N(\mu_s) =   9\varrho_B \Big(\frac{\partial\varrho_B}{\partial \mu_N}\Big)^{-1}\simeq
3 \frac{p_F^2}{\mu_N} = \frac{3 p_F^2}{N_c \mu_s} ,
\ee
whereas for quark matter,
\be
K_Q(\mu_s)  =  N_c\varrho_B \Big(\frac{\partial\varrho_B}{\partial \mu}\Big)^{-1}\simeq
\frac{N_c p_F^2}{3 \mu_s} ,
\ee
they match each other for $N_c = 3$.
One could do things even better. If the coefficient for hadron matter we redefine $9 \rightarrow 3 N_c$
and we introduce the coefficient $3/N_c$ for quark matter then both definitions match for any $N_c$ to the leading
order. At least one then could succeed in their matching around normal density.

Going back to quark matter description it would mean that
\be
K (\mu_s) =
\frac{p_F^2 (\tilde\mu_s)}{\Big(\tilde\mu_s-\sigma_1\partial_{\tilde\mu} \sigma_1 (\tilde\mu_s)\Big)}
+  3 N_c  G_\omega \varrho_B (\mu_s) ,
\ee
has a finite limit at large $N_c$ coinciding with what we get for hadron matter if one remembers that
$G_\omega \sim 1/N_c$ .

\end{document}